\documentstyle[12pt]{article}
\newcommand{\eg}{{\sl e.g.}}
\newcommand{\ie}{{\sl i.e.}}

\newcommand{\etal}{{\sl et al.}}

\input{psfig}

\begin{document}
\noindent
{\Large\bf The nature of the optical---radio correlations for powerful radio galaxies}
\vglue 1.0cm\noindent
{\bf C.N. Tadhunter$^{1}$, R. Morganti$^{2,3}$, A. Robinson$^{4}$,
R. Dickson$^{1}$, M. Villar-Martin$^{1}$ \&
R.A.E. Fosbury$^{5}$} \\
\vglue 0.5cm\noindent
{\small $^{1}$ Department of Physics, University of Sheffield, Sheffield S3 7RH \\
$^{2}$ Australia Telescope National Facility, CSIRO, PO Box 76,
Epping, NSW2121, Australia \\
$^{3}$ Instituto di Radioastronomia, CNR, via Gobetti 101, I-40129 Bologna,
Italy\\
$^{4}$ Division of Physical Sciences, University of Hertfordshire, 
College Lane, Hatfield, Herts AL10 9AB \\
$^{5}$ ST-ECF, European Southern Observatory, Karl-Schwarzschild-Strasse 2,
D85748 Garching bei Munchen, Germany} 
\vglue 1.0cm\noindent
{\bf Abstract}
The nature of the optical--radio correlations for powerful
radio galaxies is investigated using
spectroscopic observations of a complete sample of southern 2Jy radio
sources. In line with previous work, we find that significant
correlations exist between the luminosities of the [OIII]$\lambda$5007,
[OII]$\lambda$3727 and H$\beta$ emission lines and the radio luminosity.
However, our observations are not easily reconciled with the idea that
these correlations are caused by the increase in the power of the
photoionizing quasar as the jet power increases, with average ISM properties
not changing appreciably with redshift or radio power:
not only do we find that the scatter in the L$_{[OIII]}$ vs. L$_{radio}$ correlation 
is significantly larger than in L$_{[OII]}$ vs. L$_{radio}$ and L$_{H\beta}$
vs. L$_{radio}$ correlations, but the 
ionization state deduced from the emission lines does not increase with radio power as predicted by the simple, constant ISM, photionization model. We
conclude that: (a) there exists a considerable range in the quasar
ionizing luminosity at a given redshift; and (b) that the mean density
of the emission line clouds is larger in the high redshift/high power
radio sources. The latter density enhancement may either be a 
consequence of the 
increased importance of jet-cloud interactions or, alternatively, 
due to a higher pressure in the confining hot ISM, in the high redshift
objects.

Apart from the
general scatter in the correlations, we
identify a distinct group of objects with [OIII]$\lambda$5007 luminosities
which are
more than order of magnitude lower than in the general population radio
galaxies at similar redshift.
These weak 
line radio galaxies (WLRG)  are likely to be sources in which the 
central ionizing
quasars are particularly feeble.    

Deep spectra show that many of the sources
in our sample are broad line radio galaxies (BLRG). The fact that the BLRG
are observed out the redshift limit of the survey, overlapping in redshift
with the quasars, argues against the idea that BLRG are
simply the low radio power counterparts of high power, 
high redshift
quasars. Either there exists a considerable range in the intrinsic 
luminosities of the broad-line AGN for a given redshift or radio power, or
the BLRG represent partially obscured quasars. The degree of scatter present
in the L$_{[OIII]}$ vs. L$_{radio}$ correlation supports the former possibility.

\section{Introduction}

Following the first optical identifications of extragalactic radio sources 
in the 1950's it became 
clear that powerful radio emission is  associated with presence of strong
optical emission lines (Baade \& Minkowski 1954, Schmidt 1965). 
This radio-optical link is now quantified  by several
studies which show formal correlations between the optical emission line
luminosity and the radio power (Baum \& Heckman 1989, Rawlings \& Saunders
1989, 1991, Morganti \etal\ 1992, Zirbel \etal\ 1995). Given 
that the emission line regions may be energised by EUV
photons emitted in the central energy-generating region, and the radio jets
have their origin in the same region, studies of the radio/optical
correlations have the potential to provide fundamental information about
the physics of the energy generating mechanisms in radio-loud active galaxies
(see Rawlings \& Saunders 1991).

The most recent studies have revealed considerable detail in the correlations.
First, there is evidence that slope of the
correlation between radio power and emission line luminosity
is shallower for low-power FRI sources than  for the more
powerful FRII sources (Morganti \etal\ 1991, Zirbel \etal\ 1995), 
with the FRI radio sources also showing generally weaker emission lines.
Second, Laing \etal\ (1994) have identified a group of galaxies which,
despite being associated with powerful FRII radio sources, have only
weak, low ionization emission lines (these were dubbed ``low-excitation''
sources). Third, Rawlings \& Saunders (1991) found that they could reduce
some of the scatter in the correlations by plotting radio jet power (rather
than radio luminosity)
against total emission line luminosity.

It has also become clear that the emission line
properties may be affected by anisotropy
and obscuration.
Notably,
radio sources with broad-line optical spectra (BLRG, quasars) have larger [OIII] luminosities than narrow-line radio galaxies of similar extended 
radio power (\eg\ Jackson \& Browne 1990). Although it was originally thought
that this difference may provide evidence against the anisotropy-based
unified schemes for powerful radio sources (\eg\ Barthel 1989), 
the observation that the [OII] luminosities of the broad- 
and narrow-line groups are similar (Hes \etal\ 1993),
has led to the suggestion that the [OIII] emission is anisotropic, with
part of the [OIII] emitted in the obscured central regions. Furthermore,
Laing \etal\ (1994) have found that the difference between the broad- and narrow-line objects becomes less significant when the ``low-excitation'' sources
are excluded from the analysis.

The correlations between optical and radio properties are generally explained
in terms of an illumination model in which the emission line gas is 
photoionized by EUV photons emitted by a central AGN, with the 
strength
of the photoionizing continuum linked to the power in the radio jet through
the physics of the central engine (\eg\ Rawlings \& Saunders 1991). The detail
in the correlations can then be due to fundamental differences
in the nature of the central engine between the different classes of 
extragalactic radio sources (\eg\ Baum, Zirbel \& O'Dea 1995).

The illumination model has several attractions. Most notably, the
AGN photoionization models are successful at explaining the relative strengths
of most of the stronger emission lines in the spectra (Robinson \etal\ 1987).
The apparent correlation between emission line luminosity and ionization state
observed in low-z radio galaxies (Saunders \etal\ 1989) is also evidence that
the larger emission line luminosities measured in
the more powerful radio sources result from stronger ionizing continua,
rather than, for example, an increase in the covering factor of the emitting
gas with radio power or redshift. Another advantage
is that, if the illuminating radiation
field is anisotropic --- as predicted by the unified schemes --- then it
is possible to explain some of the other features of the sources, such
as the tendency for the extended emission line regions to align along the radio
axes (\eg\ Baum \& Heckman 1989) and the large UV polarization measured in some cases (Tadhunter \etal\ 1992). Finally,  
Heckman \etal\ (1994) have found a 
correlation between the radio and far-IR luminosities of 
3CR galaxies, which  provides independent evidence that the quasar luminosity 
is linked to the radio jet power.

However, the illumination model cannot explain all of the emission line
properties of powerful radio galaxies.
The highly collimated optical/UV structures which are closely aligned
with the radio axes of many high redshift radio galaxies
(\eg\ McCarthy \etal\ 1996; Best, Rottgering \& Longair 1996)
are inconsistent with the illumination of a uniform halo of
clouds by the broad radiation cones predicted
by the unified schemes. Moreover, the extreme emission line kinematics observed
in such sources
are difficult to explain in terms of AGN illumination of the {\it undisturbed} ISM of
the host galaxies (\eg\ Tadhunter 1991; McCarthy \etal\ 1996). While these
results do not rule out AGN illumination, they demonstrate that other mechanisms must also contribute.

Recent theoretical work  has  raised
the possibility of a direct link between the radio plasma and the emission-line material. Sutherland \etal\ (1993)
have proposed that the extended emission line filaments in Centaurus A are 
ionized by shocks
generated in the turbulent boundary layer between the radio jets and the
ISM. By tuning the models they can obtain a good fit to the general 
emission line 
spectra, and, crucially, their models
provide a better fit the [OIII](5007+4959)/4363 diagnostic ratio
than the single slab photoionization models. This work underlines the fact that
it can be difficult to distinguish between the emission line spectra
produced by fast shocks and AGN photoionization, because
much of the ionizing
effect of a fast shocks is due to photoionization by hot post-shock gas.
Although the application of Sutherland \etal\ model to the specific case
of Centaurus A is controversial, since the filaments lie well outside the
active inner radio jets, detailed spectroscopic
studies of other jet-cloud interaction candidates demonstrate
that the shock models may be more generally applicable
(\eg\ Clark 1996, Clark \etal\ 1997).  
The balance between
jet-induced shocks and AGN illumination is clearly a key issue for our general
understanding of the optical-radio correlations in radio galaxies.

In this paper we use spectroscopic observations of a complete sample of
southern 2Jy radio galaxies to investigate the nature of the optical--radio
correlations for powerful radio sources. In particular, we address the
question of whether the correlations can be reconciled with
the AGN illumination model. 

\section{Sample selection and data}

As originally defined by Tadhunter \etal\ (1993), our complete sample consists
of all sources from the 2Jy sample of Wall \& Peacock (1985) with declinations
$\delta < +10^{\circ}$ and redshifts $z < 0.7$. Optical, radio and X-ray observations for the complete sample are presented in Tadhunter \etal\ (1993),
Morganti \etal\ (1993) and Siebert \etal\ (1996) respectively, while new 
optical  identifications made in the course of the survey are discussed 
in di Serego Alighieri \etal\ (1994). The main advantages of this sample 
are that
it contains a good mix of objects, it covers a wide range in redshift, and,
for all the objects with redshifts $z > 0.1$,  we have complete information
about the emission line luminosities and ionization state of the emission
line gas.

Since the publication of preliminary results by Tadhunter \etal\ (1993),
deeper spectra have been obtained for most of the radio galaxies
with redshifts $z > 0.15$. As well as improving the
quality and completeness of the emission line data, these new observations 
have resulted in the detection of weak broad lines
in some of the sources previously classified as narrow line radio galaxies
(NLRG). Such sources are now re-classified as 
broad line radio galaxies (BLRG). 

The identification status of the full 2Jy sample of Wall and Peacock (1985)
has also improved since the earlier paper. Many of the new identifications
were discussed in di Serego Alighieri \etal\ (1994). The only important change
involving the $z < 0.7$ sample 
is that broad line radio galaxy 0347+05 is now included. This object was originally thought to be at higher redshifts, but spectra presented by
di Serego Alighieri et al. (1994) and Allington-Smith \etal\ (1993) place
it at a redshift of $z = 0.339$. Otherwise, all the recent identification work
has reinforced the conclusion of Tadhunter \etal\ (1993), that the 
completeness of the $z < 0.7$ sample is high; the few remaining objects in the 
full 2Jy sample without
spectroscopic redshifts are all very faint and are extremely unlikely
to fall at $z < 0.7$.

Because of its
high selection frequency (2.7GHz), the Tadhunter \etal\ (1993)
sample contains a number of
objects dominated by flat spectrum core radio
emission. However, in order to 
avoiding possible orientation biases, we require a sample 
selected on the basis of the extended, steep spectrum radio emission.
To get around this problem we have re-selected our sample as follows: the 2.7GHz core flux was
first estimated from the 5GHz core flux by assuming a flat spectra index for the
core component; the estimated 2.7GHz core flux was then subtracted from
the total 2.7GHz flux, to find the extended flux; finally, a new sub-sample was
selected which 
consists of objects with extended 2.7GHz flux greater than 2Jy. We will
refer to the new sample selected in this way as the steep spectrum selected
(SSS) subsample. As expected, 
the main effect of the reselection is to remove the majority of the
quasars dominated by compact flat spectrum core components. With the
exception of the quasar 3C273 --- which would be selected at 2Jy on the basis
of its steep spectrum jet emission alone --- all of the objects in the
SSS subsample are dominated by emission from regions with a steep high
frequency radio spectrum. 
The reselection process is not perfect, 
because the core components may be
variable, and the core spectra not exactly flat. However, we believe that
it is adequate for the purposes of the discussion below.

Basic data for the SSS subsample, including updated spectral classifications
and the latest emission line measurements, are presented in Table 1. The
radio luminosities have been calculated by
using the known power-law spectral index (2.7 --- 5GHz) to integrate
the radio spectrum over the frequency range
0.1 --- 100 GHz\footnote{$H_0 = 50$ km s$^{-1}$ Mpc$^{-1}$, $q_0 = 0.0$ are
assumed throughout this paper}. 

The classification of objects in Table 1 warrants a further discussion.
As well as classifying the objects on the basis of the
radio emission, we also classify them according to their optical properties
into five types: narrow line radio galaxies (NLRG), broad line radio galaxies
(BLRG), quasars (Q), BL Lac objects (BL), and weak line radio galaxies (WLRG). Objects in the latter class are dominated by the absorption features of the
stellar populations in the host galaxies, with small [OIII]$\lambda$5007
emission line equivalent widths ($EW([OIII]) < 10$\AA). The dividing line 
between BLRG and quasars is fuzzy: both types show broad permitted lines
in their optical spectra, but the quasars were originally classified as such
by Wall \& Peacock (1985) on the basis of their stellar appearance on optical
images, whereas the BLRG were originally classified as galaxies because
they appeared more diffuse. However, in our sample the absolute magnitude distributions of quasars and BLRG overlap. If we adopt the absolute
magnitude criterion of Veron-Cetty \& Veron (1993) for the division
between quasars and active galaxies (quasars are those objects
with $M_v < -23.0$), then PKS1547-79 and PKS2135-20
would be re-classified as quasars, while PKS1151-34 would be re-classified
as a BLRG. In Table 1 we stick with the original morphological classification
of Wall \& Peacock (1985).

The
data in Table 1 will form the basis of much of the subsequent discussion
and analysis.

\section{Results}
\subsection{Radio-optical correlations for the full sample}

Figure 1 shows the correlations between the [OIII]$\lambda$5007 
and radio luminosities for the full $z < 0.7$ sample of Tadhunter \etal\ (1993), including the latest emission-line data. In
order to examine the possible dependence of the correlations on the
radio morphology, the various radio types are indicated by different symbols.

As previously noted by Baum \& Heckman (1989),  the relationship 
between the emission line and radio luminosities is not linear. 
Interesting features include the following:
\begin{itemize}
\item{\bf FRI radio sources.} The FRI sources are invariably classified
as WLRG; they comprise the majority of upper limits in the lower left hand
part of the diagram; none have strong enough emission lines to qualify
as NLRG. Because of the large proportion of [OIII] upper limits amongst the
FRIs, it is not possible to determine from our data whether the FRIs follow a
separate correlation with a shallower slope on this diagram
(\eg\ Zirbel \etal\ 1995), or whether they merely form a continuation of 
the sequence for more powerful radio sources.
\item{\bf FRII radio sources.} While the FRII sources may follow a linear 
correlation at the high luminosity end of the diagram, below a radio luminosity
of approximately $10^{44}$ erg s$^{-1}$, the points scatter towards the
zone occupied by the FRI sources at low [OIII] luminosities. It is clear
that there exists a significant population of FRII sources with low
emission line luminosities (see also Laing \etal\ 1994). 
\item{\bf Compact steep spectrum radio sources.} If we define compact radio
sources to be those with diameters $D < 15$kpc, we find that the compact
sources with steep high frequency radio spectra 
fall on the main correlation between emission line luminosity
and radio luminosity; their emission line properties do not appear significantly
different from those of more extended radio sources of similar power. 
The emission
line properties of the compact steep spectrum radio sources in the sample
are discussed in more detail
by Morganti \etal\ 1997.
\end{itemize}  

\subsection{The correlations for different emission lines}

The incompleteness of the emission line data
for the low redshift
objects  makes further quantitative
investigation of the correlations difficult using
the full Tadhunter \etal\ 1993 sample. However, beyond
a redshift of $z > 0.1$, estimates of the 
[OIII]$\lambda$5007 and [OII]$\lambda$3727 luminosities exist for 
all quasars and
galaxies, and estimates of the H$\beta$ (narrow) luminosities are
also available for most of the galaxies (but not the
quasars). Thus, by restricting the redshift range to $0.1 < z < 0.7$, we can examine whether correlations are the same for all the emission lines.

Figures 2a,b,c show respectively  the [OIII]$\lambda$5007, [OII]$\lambda$3727 and H$\beta$ luminosities plotted against extended 
radio luminosity for all the objects in the SSS sub-sample with $z > 0.1$.
Note that the scatter in these plots is real, and not simply a consequence of
observational error: Tadhunter \etal\ (1993) estimated that the errors
in the emission line luminosities amounted to less than a factor of
two (or +/- 0.3 in the log), even allowing for possible slit losses, 
and this is borne out by
more recent repeat measurements of many of the higher redshift objects in
the sample.

It is clear that, although the degree of scatter
is similar for L$_{[OII]}$ vs. L$_{radio}$ and L$_{H\beta}$ vs.
L$_{radio}$ plots, there is a significantly larger scatter in
the L$_{[OIII]}$ vs. L$_{radio}$ plot. A particularly noteworthy
feature is that there exists a distinct group of 
four objects with [OIII] luminosities which are more than a factor of 
ten lower than would
be expected from the main correlation. This low luminosity group
comprises 0043-42, 0347+05, 1648+05(Her A) and 2211-17(3C444).
All of these objects have a low ionization state, and
with the exception of 1648+05, which has a radio
morphology somewhere between an FRI and an FRII, all the objects have an FRII radio morphology. A further interesting feature is that, despite the low {\em narrow}
emission luminosities, one of the objects in the low luminosity group is
classified as  BLRG (0347+05).
Presumably these low [OIII] luminosity objects form
part of the class of  ``low excitation'' objects identified
in the 3C sample by Laing \etal\
1994, although we prefer to label them weak line 
radio galaxies 
(WLRG) on the grounds that the meaning of the term ``excitation''
in this context is not sufficiently clear. 
The nature of the WLRGs will be discussed in more detail in
section 4.3.

Careful comparison of Figures 2a,b,c reveals that,
even if the WLRG are excluded, {\it the scatter remains
larger in the L$_{[OIII]}$ vs. L$_{radio}$ correlation than in the
correlations involving the other two emission lines}.
  
To test the significance of the correlations, we have calculated the
Spearman $\rho$ correlation coefficient for various correlations involving
the three emission lines\footnote{Note that, due to the strong correlation
between radio power and redshift in this radio flux limited sample,
it is not possible to determine whether it is the radio power or the 
redshift that is the fundamental variable. For this reason the
terms radio power and redshift are often interchangable in the text}. The results are shown in Table 2, 
from which it is clear that the null hypothesis
that no correlation exists between emission line and radio luminosities
can be rejected at a high level of significance for the  L$_{[OII]}$ vs. L$_{radio}$ correlation. However, as expected, the level of significance is lower for the L$_{[OIII]}$ vs. L$_{radio}$ correlation. 

We have also used a linear regression analysis to calculate the
slopes of the correlations assuming a linear relationship
between the two variables (on log-log plots). We find the following
relationships: 
$L_{[OIII]} \propto (L_{radio})^{0.308\pm0.015}$ and
$L_{[OII]} \propto (L_{radio})^{0.576\pm0.010}$. 
Note that in each case we have assumed that the emission line
luminosity is the dependent variable. The
WLRG have been excluded from this analysis, since they 
form a distinct group of objects which falls off the main
correlations.
  
\subsection{Ionization state and luminosity}

Investigations of the relationship between ionization state and
luminosity provide 
important information about the nature of the optical--radio correlations
for powerful radio galaxies.

Previous results presented by Saunders \etal\ 1989 for a complete
sample of low redshift 
radio galaxies  ($z < 0.2$) showed evidence for an anti-correlation
between the
ionization-sensitive
[OII](3727)/[OIII(5007) ratio and L$_{[OIII]}$, implying that the more luminous 
objects have a higher ionization state. Our data now allow us to
investigate the ionization/luminosity relationship in greater
depth and, in particular, to determine whether the trends noted by
Saunders \etal\ continue to higher redshifts and higher radio powers.

Figures 3a,b,c and d show   
[OII](3727)/[OIII](5007)
plotted against L$_{[OII]}$, L$_{[OIII]}$, L$_{H\beta}$ and L$_{rad}$
repectively for the $z > 0.1$ SSS sub-sample. The Spearman's $\rho$
correlation coefficients and the associated significance levels
are listed in Table 2. For comparison,
the data for objects in the Saunders \etal\ (1989) sample and assorted
higher redshift 3C galaxies from the literature are plotted in Figure 4.

Concentrating first on [OII]/[OIII] vs. L$_{[OIII]}$ (Figure 3a), we find that
we can confirm the  Saunders \etal\ result: for the lower
redshift ($0.1 < z < 0.2$) objects in our sample there is a definite
trend of decreasing [OII]/[OIII] with [OIII] emission line luminosity. However,
considering the full redshift range of our sample ($0.1 < z < 0.7$), the
trend is much weaker. In fact, there are signs that the slope of the relationship flattens off at high emission line luminosities, especially
if the WLRG are removed. Part of the reason for the weakness of the
trends for the full redshift range becomes clear when the high ($0.2 < z
< 0.7$)
and low ($0.1 < z < 0.2$) redshift objects are considered separately: with
the exception of the WLRG 0347+05, the high
and low redshift objects occupy separate regions of the diagram, with
the high reshift objects appearing to form a parallel
sequence to the low redshift
objects. {\it It is clear that 
high redshift objects do not form a 
high ionization continuation of the correlation found
for the lower redshift objects}. Similar trends are noted in Figure 4
for the 3C objects (see also Jackson \& Rawlings 1997), although 
in this case the sample is not complete beyond
a redshift of $z = 0.2$.

At this point it should be stressed that some care is
required when interpreting the [OII]/[OIII] vs. L$_{[OIII]}$ anti-correlation, 
because of the mutual
dependence of both  variables on the [OIII] luminosity.
For example, even for the case in which the 
L$_{[OII]}$ and L$_{[OIII]}$ luminosities
are uncorrelated, and there is no genuine physical correlation between
ionization state and luminosity, a spurious anti-correlation would nonetheless
be observed simply because of this effect. In the case of the data plotted in Figure 3, we note that
there is in fact a strong correlation between L$_{[OIII]}$ amd L$_{[OII]}$
(Table 2), and
this effect is unlikely to be the sole cause of the anti-correlation
observed for the low redshift objects in Figure 3a. 

It is significant that, despite
the possible anti-correlation between the [OII]/[OIII] 
ratio and [OIII]$\lambda$5007
luminosity observed at low redshifts, no
significant anti-correlations are found
in the plots involving the other emission lines and the radio luminosity
(Table 2, Figures 3b,c,d).
In the none of these diagrams do the high power/high redshift sources
form a continuation of the sequence defined by the low power/low redshift
source. This result holds for both the broad and the narrow line
objects, which are indicated by different symbols in the diagram. 

On balance, the evidence for a strong anti-correlation between
ionization state and luminosity
is weak when we consider the full redshift range of the sample. Table 3 shows
the median of the measured values of [OII](3727)/[OIII](5007) 
ratio for different
redshift ranges. It is clear from this that the 
median [OII]/[OIII] does not vary
significantly between $0.1 < z < 0.2$ and $0.5 < z < 0.7$ despite more than
order of magnitude increases in both the median radio luminosity and 
the median emission line luminosity. Indeed, the highest redshift and most
luminous radio source in the sample (0409-75) has one of the lowest ionization
states as indicated by the [OII]/[OIII] ratio; recent observations
of high redshift 3C radio galaxies also show that many of them have relatively
low ionization states (Jackson \& Rawlings 1997). 
The implications of these results for understanding
the nature of the optical--radio correlations will be discussed in
section 4.1.

\subsection{Broad line objects}

In common with the recent work of Laing \etal\ (1994), we find that many of the
objects in the sample prove to be BLRG when observed at sufficiently high
S/N. The fact that none of these BLRG is highly polarized at
UV wavelengths (see Tadhunter \etal\ 1997) indicates that
we are seeing the AGN directly in these objects rather than via scattering in the ISM of the host galaxies. 

We detect BLRG out to the redshift
limit of our sample. This shows that BLRG are not merely the low redshift/low
radio power counterparts of high redshift/high power quasars; there is
an overlap in radio power ranges covered by the populations of BLRG
and quasars. This implies that, either there exists a considerable
range in the intrinsic luminosities of broad line AGN at a given redshift/radio
power, or the range of intrinsic luminosities is small but there
is a range in the extinction, with the BLRG representing
partially obscured quasars. The obscuration idea is supported by spectroscopic 
observations of low redshift BLRG which show large  Balmer 
decrements as well as non-stellar continua which are red when compared
with quasar continua (\eg\ Osterbrock, Koski
\& Phillips 1976). The relatively strong Paschen lines measured
in some BLRG may also be consistent with large reddening
(\eg\ Rudy \& Tokunaga 1983, Lacy \etal\ 1982), although 
it remains a possibility in some cases that the large 
Balmer decrements are due to collisional excitation
effects in the BLR clouds (\eg\ Kwan \& Krolik 1981), and 
that the BLRG continua 
are {\it intrinsically} red. 

Concentrating on the 32 objects in the SSS sub-sample with $0.1 < z < 0.7$,
we find that 14 (44$\pm$14\%) are broad line objects
(\ie\ BLRG or quasars). Dividing the sample by redshift, the occurrence
rate of broad line objects is 54$\pm$27\% in the
high redshift half of the sample ($0.4 < z < 0.7$), and 38$\pm$16\% 
in the low redshift half of the sample ($0.1 < z < 0.4$). Although
the relative frequency of broad line objects appears to increase with redshift,
the trend in not statistically significant, given the relatively small numbers
of objects in the samples. These results are entirely consistent with 
those obtained by Lawrence (1991) for 3CR objects in the same redshift range.

It is also important to consider whether the populations
of broad- and narrow line objects have
significantly different [OIII]$\lambda$5007 emission-line luminosities, as
has been found for other samples (\eg\ Jackson \& Browne 1990). 
Figure 5 shows L$_{[OIII]}$ vs. 
L$_{rad}$ for the SSS sub-sample, with broad- and narrow-line objects
indicated by different symbols. While there may be some weak evidence
from this diagram for the broad line objects to show higher [OIII]$\lambda$5007
luminosities, this result is not as strong 
as found in some previous studies. Using a Kolmogorov-Smirnov two sample
test we find that the significance of the difference in the distributions
of [OIII]$\lambda$5007 luminosities for broad- and narrow-line 
objects in the SSS sub-sample
is only 10\%. In line with the results of Laing \etal\ (1994), the 
differences
become even less significant if the WLRG are removed from the analysis.

\section{Discussion}

\subsection{Testing the AGN illumination model}

It is generally assumed that the emission line properties of the
powerful radio galaxies can be explained in terms of a quasar illumination
model. 
 
In what we will refer to as the basic quasar illumination model, the emission
line regions are assumed to be photoionized by EUV photons from a 
central illuminating
quasar or AGN; and 
the properties of the ISM (e.g. density, covering factor etc..)
are assumed not to vary on average with redshift or radio power, although
there may be a scatter in the ISM properties at a given redshift.
In this model the correlations between emission line 
luminosity and radio power arise because 
of the photoionizing luminosity of the illuminating quasar and  
the radio jet power are tied through the
physics of the central engine. 
We now examine whether  this basic model be reconciled 
with the spectroscopic
data.

As a starting point we consider the behaviour of the emission line flux and line
ratios for a single, low density, optically thick slab of solar 
abundances, as the
strength of the ionizing continuum is increased. The model results were
generated using the MAPPINGS photoionization code for a power-law ionizing 
continuum shape ($F_{\nu} \propto \nu^{+\alpha}$), and
are quantified as sequences in the ionization parameter
defined by: $U = F_{ion}/n_ec$, where $F_{ion}$ is the 
number of ionizing photons per unit area at the face of the slab, 
$n_e$ is the electron density
in the cloud, and $c$ is the speed of light.
The models make the
following general predictions:
\begin{itemize}
\item{\bf Variation in [OIII]$\lambda$5007 luminosity with ionizing continuum
luminosity.} Over the range of ionization parameter consistent with
the emission line ratios measured in low redshift radio galaxies ($10^{-3} < U < 10^{-1}$:
Robinson \etal\ 1987) the [OIII] luminosity rises steeply with the ionizing
continuum luminosity ($L_{[OIII]} \propto (L_{ion})^{+\beta}$
with $\beta = 1.2$ for $\alpha = -1.5$). This result holds for a wide range of ionizing continuum
shapes ($1.1 < \beta < 1.3$ for $-1.0 > \alpha > -2.0$).
\item{\bf Variation in [OII]$\lambda$3727 luminosity with ionizing continuum luminosity.}
Over a similar range of ionization parameter, the [OII]$\lambda$3727
luminosity rises much
less steeply with ionizing continuum luminosity than the 
[OIII]$\lambda$5007 luminosity
($L_{[OII]} \propto (L_{ion})^{+\beta}$ with $\beta = 0.46$ for 
$\alpha = -1.5$). 
In this case there is a larger variation in the slope of the relationship 
with the shape of
the ionizing continuum ($0.8 < \beta < 0.4$ for $-1.0 > \alpha > -2.0$), 
but the 
slope is always predicited to be less than in the case of the relationships
between $L_{[OIII]}$ and $L_{ion}$
\item{\bf Variation of ionization state with ionizing continuum luminosity.}
The ionization state
is predicted to increase as the ionizing continuum luminosity is increased
at fixed $n_e$. In
consequence,
the [OII](3727)/[OIII](5007) ionization diagnostic ratio will decrease as 
both the [OIII] and [OII] emission line luminosities increase, but
the rate of decrease of [OII]/[OIII] is predicted to be significantly
stepper with [OII] than with [OIII] ($[OII]/[OIII] \propto (L_{[OIII]})^{-0.64}$
and $[OII]/[OIII] \propto (L_{[OII]})^{-1.7}$ for an $\alpha = -1.5$
ionizing continuum). 
\item{\bf Diagnostic diagrams.} The models form unique
sequences in the ionization
parameter on the diagnostic diagrams, with the position of each 
sequence depending on the shape of the ionizing continuum
(\eg\ Robinson \etal\ 1987).
\end{itemize}
We emphasise that these results were derived from single slab models. In
reality, our spectroscopic observations will encompass an ensemble of
emission line clouds in each radio galaxy, covering a wide range of radii and density. 
However, provided that the condition of the clouds being optically
thick holds, and the densities do not become so high that
collisional de-excitation effects become important, then we expect these predictions to apply to
ensembles of clouds: the ensemble as  a whole will move up and down the
ionization parameter sequences as the ionizing continuum luminosity is varied.

At least some features of our observations are consistent with the single
slab photoionization models. Figure 6 shows the 
[OII](3727)/[OIII](5007) vs. [OIII](5007)/H$\beta$ diagnostic 
diagram with data points from 
the $z > 0.1$ SSS subsample compared with the photoionization model predictions
for various photoionizing continuum shapes. With the exception
of some of the extreme low ionization points, the data points
form a tight sequence on this diagram, with a power-law 
photoionization model providing the best fit to the data for a spectral
index $\alpha = -1.5$. This is entirely consistent with the results obtained 
for a sample of low
redshift radio galaxies by Robinson \etal\ (1987), despite the fact that
the current sample extends to much higher redshifts and radio powers. Note, 
however, that this agreement with the photoionization model predictions 
does not rule out other ionization mechanisms: the predictions of fast
shock models overlap with the predictions of the photoionization models in
some sectors of the diagnostic diagrams (\eg\ Sutherland \etal\ 1993). 

The photoionization models also provide a ready explanation for
the increased scatter in the L$_{[OIII]}$ vs. L$_{radio}$ correlation
relative to the  correlations involving the other emission lines. 
If the scatter in the emission
line luminosity reflects a range in intrinsic quasar ionizing luminosities
at a given radio power, then the photoionization models predict a greater
scatter in the L$_{[OIII]}$ vs. L$_{radio}$ diagram, 
because the [OIII] emission lines
are more sensitive to variations in the ionizing continuum than the [OII] lines
over the relevant
range of ionization parameter. This effect also explains why the 
WLRGs fall well below the locus of
points for the other radio galaxies in the L$_{[OIII]}$ vs. L$_{radio}$ 
diagram, but
fall much closer to the locus in the L$_{[OII]}$ vs. L$_{radio}$ diagram.
Note that the data allow us to rule out the possibility
that the scatter in the 
correlations 
is solely due to a variation in the covering factor of
the emission line clouds, since a variation of the covering factor 
would affect [OIII]$\lambda$5007,
[OII]$\lambda$3727 and H$\beta$ equally, and the scatter would be the same
in all the correlations.  

The evidence to support the quasar illumination model based on the slopes of
the correlations between radio power and emission line luminosity is more
ambiguous.  Although all the emission line luminosities increase with radio
luminosity, as expected in the case of a direct link between the luminosity in
the quasar ionizing continuum and the radio jet power, the [OII] luminosity
appears to increase more rapidly with the radio luminosity than the [OIII]
luminosity (section 3.2, Figure 2), in apparent contradiction of the
photoionization model results.  However, given the large scatter and
relatively low significance of the L$_{[OIII]}$ vs.  L$_{radio}$ correlation,
this does not by itself represent strong evidence against the quasar
illumination model.  Furthermore, the slope of the L$_{[OII]}$ vs. 
L$_{radio}$ correlation is not far from the prediction of the quasar
illumination model. 

More worrying for the basic illumination model is the lack of evidence for
a strong
anti-correlation between the ionization-sensitive [OII](3727)/[OIII](5007) 
ratio and the emission line
luminosity (see section 3.3). 
On the basis of the single slab model, we would expect the high power
objects at high redshifts to form a high ionization extension to any
correlation found for the low redshift sources, but in fact the high redshift
sources form a parallel locus in the diagrams, and cover a similar range in 
ionization to the low redshift sources (Table 3).

Thus, although the basic quasar
illumination model is successful in some aspects, 
it cannot explain all the features of our data. 
We now consider the alternatives.

\subsection{Modifications to the quasar illumination model}

We emphasise that failure of the quasar illumination model to explain all
features of the data does not imply that the general idea of photoionization
of the emission line clouds by the quasar nucleus should be abandoned.  If the
unified schemes for radio-loud active galaxies are correct, then all powerful
radio galaxies will harbour luminous quasars in their cores which, for typical
ISM properties, will have a significant effect on the ionization of the
emission line clouds.  Quasar illumination is also supported by the detection
of scattered UV continuum emission and broad lines in the extended
nebulosities of many powerful radio galaxies (\eg\ Cimatti \etal\ 1995, Young
\etal\ 1996), including some objects from the SSS (Tadhunter \etal\ 1997);
such polarized features cannot not be produced in any other way than quasar
illumination. 

Rather we should look to the assumptions of the simple illumination model, and 
consider what modifications are required to bring the illumination model into
agreement with the observations.

First, we consider the assumptions of the single slab photoionization models. 
In the above discussion it was assumed that the photoionized clouds are large
enough to be optically thick to the photoionizing continuum (\ie\ radiation
bounded).  Relaxation of this assumption could lead to a decrease in the
strength of [OII]$\lambda$3727 relative to [OIII]$\lambda$5007, because
optically thin clouds have less extensive partially ionized zones in which low
ionization lines such as [OII]$\lambda$3727 are produced.  Thus, in order to
explain the relatively strong [OII] emission in the high redshift objects we
require that the clouds become more optically thick as the radio power and
redshift are increased.  However, if the ISM properties are the same on
average at all radio powers and redshifts, then it is inevitable that, as the
quasar ionizing continuum luminosity is increased we will observe {\it more}
optically thin clouds in the higher power sources --- the reverse of what we
would require to explain the stronger [OII] emission by this mechanism. 

A major assumption of the simple quasar illumination model is that, on
average, the properties of the ISM do not change as a function of redshift or
radio power. A plausible explanation for the relatively strong [OII], and
high [OII]/[OIII] ratios observed in the high redshift objects is that, 
in reality, there is significant evolution in the
density distribution of the clouds as a function of redshift or radio
power. For a given ionizing luminosity and radius from the ionizing source,
the ionization state of the gas is inversely proportional to the density. Thus,
an increase in the mean density of the emitting clouds with radio power or 
redshift could compensate for the increase in the strength of the ionizing
continuum, and ensure that the range of ionization observed in high and low
redshift objects is similar. Such a change in the density is plausible and might arise in two ways:
\begin{itemize}
\item{\bf Evolution in the environment and confining medium.} It is generally
assumed that the warm emission line clouds are confined by a hot X-ray emitting
phase of the ISM associated with the host galaxy (\eg\ Forman, Jones \& Tucker
1985) or cluster of galaxies (\eg\ Fabian, Nulsen \& Canizares 1984). 
In this case, the 
density of the warm clouds will be related to the pressure of the hot confining
medium, which will be  larger close to the centres of rich clusters of
galaxies than close to the centres of isolated galaxies. Thus, it is plausible
that the high redshift/high power sources have larger gas densities because they are in
richer environments than their low redshift/low power counterparts. Indeed, 
optical imaging studies 
of the fields of radio galaxies provide strong evidence for the required evolution
in the radio source environments (Hill \&  Lilly 1991), 
while X-ray observations of at least some
powerful high redshift radio galaxies show direct evidence for a high pressure IGM
(Crawford \& Fabian 1994,1996).  The fact that the nearby powerful
radio galaxy Cygnus A --- one of the few nearby radio galxies of comparable power
to the high redshift radio galaxies --- also lies at the heart of a rich cluster
with a dense IGM, suggests that the evolution in the environments is an evolution
with radio power rather than redshift. It is also notable that the relatively large
[OII]/[OIII] ratios measured in extended emission line regions around radio-loud quasars have
been used to {\it deduce} the presence of a high pressure confining medium at high redshifts (Crawford \& Fabian 1989, Forbes \etal\ 1990).
  
\item{\bf The compression effect of jet-cloud interactions.} Gas along the radio axes
will be compressed as it passes through the shocks driven through the ISM by the
radio jets and enters the jet coccoon (the gas may
also be ionized as it passes through the shocks --- see below), 
with typical compression factors of order x100 or more. Hence, if the jet-cloud interactions
become more important in the high redshift sources, the densities will be higher, the
ionization state lower, and the [OII]$\lambda$3727 relatively stronger. This is before considering
the additional ionizing effects of the jet shocks (see below). There is now plenty of
evidence to support the idea that the jet-cloud interactions become relatively
more important as the redshift increases: not only do the extended structures 
become more closely
aligned with the radio axes (\eg\ McCarthy \etal\ 1987), but the emission line kinematics become more extreme as
the redshift increases (McCarthy \etal\ 1996). 
Furthermore, accurate measurements of densities and pressures
in the EELR along the radio axes of nearby radio galaxies provide direct evidence for
pressures significantly greater than indicated by X-ray observations of the hot ISM
in the same galaxies (Clark 1996, Clark \& Tadhunter 1996).  Recently Best \etal\ (1997) have found evidence for a correlation between the optical/UV morphologies of the high redshift radio galaxies and
the diameters of the radio sources, in the sense that the objects with the highest surface brightness, most closely aligned structures have the
most compact radio sources. In the context of the jet-cloud interaction
model these results can be explained in terms of an increasing
dominance of the jet-induced shocks in the more compact sources. Indeed it
is notable that two of the high redshift sources in Best \etal\
with compact radio sources and closely aligned radio and optical structures
--- 3C368 and 3C324 --- have relatively low ionization emission line spectra
([OII](3727)/[OIII](5007)$\sim$1 Jackson \& Rawlings 1997), consistent with
the dominance of jet-induced shocks, whereas the one object with a broad, roughly
bi-conical emission line distribution and relatively extended radio source --- 3C265 ---
has a high ionization emission line spectrum, consistent with the dominance of AGN
illumination of the ambient ISM (see Figure 4). It will be important in the future to investigate in more detail the links between the alignment effect, the compactness of
the radio source and the ionization state of the emission line gas.
\end{itemize}
An alternative possibility is that, due to
the particular combination of the radial density law and geometrical
dilution of the radiation field, the
ionization state of the gas decreases with radius.
Then, if the mean radial distance of the emitting clouds from the nucleus is larger
at high redshifts, lower ionization states will result than in the
constant radius situation. Such a situation might arise,
for example, if the warm gas originates in mergers, but has yet to settle into
a stable configuration in the central regions of the host galaxies of the high redshift
sources. Alternatively, the emission line clouds at
small radii
could be more heavily obscured by the central dust lane in the
higher redshift sources. The published data on the emission line structures of
powerful radio galaxies do show some evidence for a change in the distribution of
emission lines with redshift: whereas the emission lines in most 
low redshift radio galaxies
show a strong, dominant peak in the emission line distribution centred on
the continuum nucleus (Tadhunter 1987, Baum \etal\ 1988), the emission lines
in high redshift radio galaxies ($z > 0.5$) often peak well away from the continuum
nucleus, with a large fraction of the emission line flux
emitted outside a radius of 10kpc (McCarthy \etal\ 1996).
However, it is not yet clear whether this change in the
emission line distribution is  another manifestation of jet-cloud interactions,
which become more important at high redshifts, or reflects a real change
in the intrinsic distribution of warm/cool interstellar medium with redshift or
radio power. 

It is clear that the observed correlations between optical emission line
and radio properties can be reconciled with the illumination model provided that
the mean density and/or radial distribution of the emitting
clouds evolve with redshift or radio power.
 
\subsection{Alternative models}

Given the close alignments between optical and radio structures and the
extreme emssion line kinematics observed in some sources, an obvious 
alternative to AGN illumination is ionization by the shocks driven through
the ISM by the radio jets.

One effect of the jet shocks will be to compress the
warm clouds. This will inevitably lead to a lower ionization state, even if
an illuminating quasar provides the bulk of the ionizing energy.
What is less clear is whether the shocks can also provide the bulk of the
ionizing energy for the clouds. If they do, then the relationship between
the radio and optical components is more direct than implied
by the simple illumination model.  

In this context it is instructive to consider the cases of low redshift
radio galaxies in which there is clear evidence for detailed associations
between extended emission line regions (EELR) and radio structures (see Clark
1996). In such cases emission line diagnostics not only provide direct
evidence for the compression effect of the jet shocks, but, in addition, certain
of the diagnostic ratios -- notably HeII(4868)/H$\beta$ and
[OIII]4363/(5007+4959) --- are more consistent with shocks 
than with photoionization. It is important to note however that the emission
line gas is roughly co-spatial with the radio structures in these objects,
whereas for this paper we are considering the the near-nuclear 
emission line regions ($r < 20$ kpc) which are often on 
a much smaller scale than the extended radio structures (with the notable
exception of the CSS sources).

Although there is little direct, morphological evidence for jet-cloud
interactions in the near-nuclear regions, this does not rule out a significant
ionizing input from the jet shocks, which could be driven tangentially through
the ISM. Detailed observations of the kpc-scale gas in the 
powerful nearby
radio galaxy Cygnus A, reveal extreme high velocity components which would
be difficult to explain in any other way than jet-cloud interactions
(Tadhunter 1991). 
Furthermore, the spatial variations in some of the emission line ratios in the
same source (notably [NII]$\lambda$6584/H$\alpha$) defy explanation in terms of
a simple AGN illumination picture (Tadhunter \etal\ 1994).

Set against the evidence for jet-cloud interactions is 
the strong evidence for AGN illumination. This includes: the polarimetric
evidence for scattered quasars in several powerful radio galaxies; the fact
that extended emission line gas in most low redshift galaxies lies well
away from the radio axis and has relatively quiescent kinematics; and
the emission line diagnostic diagrams. Note that, although the
success of the photoionization models in the diagnostic diagrams
does not rule out other ionization mechanisms --- the results of
the shock and photoionization models overlap in some regions of the diagnostic
diagrams --- pure photoionization models can explain the {\it sequences} 
on these
diagrams more naturally than pure shock models.

If we were to abandon AGN photoionization entirely, it would then be 
difficult to explain the  good correlations between radio and optical 
emission line luminosities
by shocks alone. Variable factors in the shock models include the 
distribution of gas relative to the radio structures, the shock speed, 
and the relative importance of precursor gas and shocked gas. While it is
true that the radio jets are likely to be more powerful in the
more luminous radio sources, it is not at all clear that this will translate
into a good correlation between emission line luminosity and radio power via
jet-cloud interactions.  

Given that the closer UV/radio alignments and the more extreme emission line
kinematics observed in the high redshift radio galaxies, it is plausible that
the jet shocks become an increasingly important source of ionizing energy as
the radio power or redshift increase.  Since the shocks can produce a
relatively low ionization state, the shock ionization might also help to
explain the lower-than-expected ionization states observed in the higher
redshift objects. 

However, it is unlikely that the jets provide the bulk of the ionizing energy
over the entire range of radio power and redshift. 

\subsection{The nature of the WLRG}

In section 3.2 we identified a distinct group of weak  line
radio galaxies (WLRG) with [OIII]$\lambda$5007 luminosities
more than a factor of ten below those of the general population of 
radio galaxies at similar redshifts. All of these WLRG also show a
low ionization state, as indicated by the [OII](3727)/[OIII](5007) ratio.
How do we explain the low [OIII] luminosities of these objects? In the
context of the quasar illumination model there are
several possibilities:
\begin{itemize}
\item[(a)]  The illuminating quasars are much 
weaker in the WLRG.
\item[(b)]  The covering factor of the emission line clouds is lower.
\item[(c)] The gas density is much higher than average, and this leads to
lower ionization state, and relatively weaker [OIII]$\lambda$5007 emission.
\item[(d)] The radio sources are over-luminous (e.g. Barthel \& Arnaud 1997).
\end{itemize}
The lower covering factors idea (possibility (b)) cannot work by 
itself because, while lower covering factors will
lead to lower emission line luminosities, this would not explain the lower ionization states observed in the WLRG.
It is also clear that higher densities (possibility (c)) cannot be the 
sole cause of the lower [OIII] luminosities, because the H$\beta$
luminosities --- which are a direct measure of the number of ionizing
photons intercepted by the clouds --- are also significantly lower than the
mean in at least two of the sources (2211-17 and 1648+05).

An alternative explanation is that emission
line clouds in WLRG have both a higher density {\it and} a lower 
covering factor than the general population of powerful radio
galaxies (\ie\ a combination of (b) and (d)). If the emission line clouds
are spherical and have the same mass distribution on average in all radio galaxies,
then such a combination could arise because the pressure in the hot confining
medium is much larger in the WLRG:  the density will rise and the covering
factor will decrease as the confining pressure increases. The lower ionization 
states measured in the WLRG
imply a density and pressure that are a factor of $\sim$10 higher than in the general population of powerful radio galaxies, for a given
AGN luminosity. Since the covering factors
of the spherical clouds scale as (pressure)$^{-2/3}$, a factor $\sim$10 increase
in density and pressure implies a factor $\sim$5 {\em decrease} in the covering
factor. Such a decrease is consistent with the lower H$\beta$
luminosities of WLRG. 

Another explanation related to the environments of the sources is
that, rather than the emission
lines being under-luminous, the radio sources are over-luminous, due to
their interaction with the dense, hot interstellar medium in rich clusters
of galaxies (possibility (d): see Barthel \& Arnaud 1996). In this
context it is notable 
that 1648+05 and 2211-17 are the
among the most luminous X-ray sources in the SSS subsample, with direct evidence
for extended, high-pressure X-ray haloes (Siebert \etal\ 1996). On
the other hand, while the X-ray observations of 0347+05 are
inconclusive, the X-ray observations of 0043-42 show no evidence for
a dense confining medium. 
In addition, at least some powerful radio galaxies in dense environments show strong, high ionization emission lines
(\eg\ Cygnus A: Tadhunter \etal\ 1994). 

Although the relationship between the emission line luminosity
and the environment requires further investigation, the simplest  
explanation for the properties 
of the WLRG is that the illuminating AGN in these sources are an order of magnitude less powerful than in the general population of radio galaxies at similar redshifts (possibility (a)).
If correct, this explanation would have important implications for the unified schemes because it would imply that not all powerful radio galaxies have luminous quasars 
in their cores.

\subsection{Implications for the unified schemes}

In a statistical sense our observations support the anisotropy-based unified schemes for powerful radio sources: the proportion of broad line objects
in the SSS is comparable to that found in samples of 3C radio galaxies; 
the implied opening angle for the quasar radiation cone 
(56$\pm$10$^{\circ}$) is consistent with estimates based on comparison of
the linear size distributions of 3C quasars and radio galaxies (Barthel 1989);
and the differences between the [OIII] luminosities of broad and narrow line
objects in our sample are small, especially when the WLRG are removed from
the analysis. In a recent analysis of the core radio emission of sources
in  the SSS subsample, 
Morganti \etal\ 1996 found that the broad line objects  
show a tendency to have relatively stronger radio cores than the narrow line
objects in the same sample, where the relative core strength ($R$) is quantified as
the ratio of core to extended radio flux at a particular frequency.
(\ie\ $R = S_{core}/S_{ext}$). Again this is consistent with the 
anisotropy-based unified schemes in which the radio core emission is strongly
anisotropic due to relativistic beaming effects.

Despite the positive {\it statistical} evidence, there are individual objects
which cannot be readily accommodated within the unified schemes. Two objects
stand out: PKS0347+05 and PKS1549-79\footnote{Note that PKS1549-79 is part
of the Tadhunter \etal\ (1993) sample, but it is excluded from the SSS sub-sample because
its radio emission is dominated by a flat spectrum core component}. 
On the basis of
unified schemes we expect the
objects with the radio axes pointing close to the line of sight 
to have optical spectra dominated by broad lines and relatively
strong radio cores (\ie\ large $R$), while the objects with radio axes
close to the sky plane should have optical spectra dominated by narrow lines
and relatively weak radio cores (small $R$). However, PKS0347+05
has extremely weak radio core ($R_{2.3GHz} < 0.0006$)  yet its optical spectrum
is dominated by broad lines with only weak narrow lines, whereas in  the
case of PKS1549-79 we find an extremely strong radio core 
($R_{2.3GHz} = 1.310$), strong
narrow lines, but no sign of broad permitted lines. In
showing exactly the reverse pattern
of behaviour to that predicted by the unified schemes,
these objects provide a timely reminder that the unified schemes, if they have
any validity, are only
valid in a statistic sense; the quasar cone opening angles, intrinsic
quasar luminosities, bulk jet powers and ISM properties are likely to vary
consideraby from source to source. The existence of a group of objects in
which the illuminating quasar emission is apparently weak or non-existent
(the WLRG), 
futher emphasises this point.

\section{Conclusions}

In line with previous results, 
we find strong, formal correlations between the optical emission line luminosity
and the radio luminosity for 
our complete sample of southern 2Jy radio sources. 
A major new result is that different emission lines show
different behaviours: the scatter in the L$_{[OIII]}$ vs. L$_{radio}$ correlation 
is greater than in either the L$_{[OII]}$ vs.  L$_{radio}$  or the L$_{H\beta}$ vs.
L$_{radio}$ correlations. There also exists a distinct group of objects (the WLRG) in
which the [OIII] luminosity is more than an order of magnitude less than in the
general population of powerful radio galaxies with similar radio power. In the context
of the quasar illumination model the most natural explanation for these results is that
there exists a considerable range of quasar luminosity for a given radio power, and
not all powerful radio galaxies contain luminous quasars.

In the simplest quasar illumination model the correlations between radio and
optical luminosities are explained in terms of the
power of the photoionizing quasar increasing
in proportion to the radio jet power, with the properties of the
ISM remaining constant (on average) as the radio power is increased.
We have made a critical test of this simple model by comparing our results
with single slab photoionization models. We find that the high 
redshift/high radio power objects do not form a high ionization extension
to the correlation between ionization and luminosity found for the low redshift radio
galaxies; the range of ionization measured in the high redshift 
($0.5 < z < 0.7$) objects
is not significantly different from that measured in the lower power,
low redshift ($0.1 < z < 0.2$) objects. These results can  be reconciled with
the simple quasar illumination model provided that  the radial distribution
and/or the mean density of emission line clouds varies with  
with radio power or redshift. Density enhancements could arise because either
the higher power objects are in richer cluster environments with a higher pressure confining
ISM, or the compression effect of shocks driven through the ISM by the radio
jets becomes increasingly important as the redshift and radio power increase. Given the
strong evidence for jet/cloud interactions in many of the high redshift radio
galaxies, we favour the latter explanation.
\vglue 0.3cm\noindent
{\bf Acknowledgments.} This work is based on observations taken using the 
European Southern Observatory 3.6m telescope,
La Silla, Chile. We thank the STARLINK project for the support of computer
facilities in Sheffield. MVM and RD acknowledge support from PPARC. RAEF
is affliated to the Astrophysics Division, Space Science Department, European
Space Agency.
\vglue 0.3cm\noindent
{\bf References}    
\begin{description}
\item Allington-Smith, J.R., Peacock, J.A., Dunlop, J.S., MNRAS, 1991,
253, 287
\item Baade, W., Minkowski, R., 1954, ApJ, 119, 215
\item Barthel, P.D., 1989, ApJ, 336, 606
\item Barthel, P.D., Arnaud, K.A., 1996, MNRAS, 283, L45
\item Baum, S.A., Heckman, T., Bridle, A.H., van Breugel, W., Miley, G., 1988,
ApJS, 68, 643 
\item Baum, S., Heckman, T., 1989, ApJ, 336, 702
\item Baum, S.A., Zirbel, E.L., O'Dea, C.P., 1995, ApJ, 451, 88 
\item Best, P., Longair, M., Rottgering, H., 1996, MNRAS, 280, L9
\item Cimatti, A., Dey, A., van Breugel, W., Antonucci, R., Spinrad, H., 1995,
ApJ, 465, 145.
\item Clark, N.E., Tadhunter, C.N., 1996. In: {\it Cygnus A -- Study of
a Radio Galaxy} (eds Carilli, C.L. \& Harris D.E.), CUP, p15
\item Clark, N.E., 1996, PhD Thesis, University of Sheffield
\item Clark, N.E., Tadhunter, C.N., Morganti, R.M., Killeen, N.E.B., Fosbury,
R.A.E., Hook, R.N., Siebert, J., Shaw, M.A., 1997, MNRAS, 286, 558
\item Crawford, C.S., Fabian, A.C., 1989, MNRAS, 239, 219
\item Crawford, C.S., Fabian, A.C., 1993, MNRAS, 260, L15
\item Crawford, C.S., Fabian, A.C., 1995, MNRAS, 273, 827
\item Dickson, R.D., 1997, PhD thesis, University of Sheffield
\item di Serego Alighieri, S., Danziger, I.J., Morganti, R., 
Tadhunter, C.N., 1994, MNRAS, 269, 998
\item Fabian, C.N., Nulsen, P.E.J., Canizares, C.R., 1984, Nature, 216, 733
\item Forbes, D.A., Crawford, C.S., Fabian, A.C., Johnstone, R.M., 1990, MNRAS, 244, 680
\item Forman, W., Jones, C., Tucker, W., 1985, ApJ, 293, 102
\item Heckman, T.M., O'Dea, C.P., Baum, S.A., Laurikainen, E., 1994, ApJ,
428, 67
\item Hes, R., Barthel, P., Fosbury, 1993, Nature, 362, 326
\item Hill, G., Lilly, S.J., 1991, ApJ, 367, 1
\item Jackson, N., Browne, I.W.A., 1990, Nature, 343, 43
\item Jackson, N., Rawlings, S., 1997. MNRAS, 286, 241
\item Kwan, J., Krolik, J.H., 1981, ApJ, 250, 478
\item Laing, R.A., Jenkins, C.R., Wall, J.V., Unger, S.W., 1994,
In: {\it 1st Stromlo Symposium on the Physics of Active Galaxies}, 
Bicknell, G.V., Dopita, M.A., Quinn, P.J. eds, ASP
Conference Series, 54, p227
\item Lacy, J.H., Soifer, B.T., Neugebauer, G., Matthews, K., Malkan, M.A., 
Becklin, E.E., Wu, C.-C., Boggess, A., Gull, T.R., 1982, ApJ, 256, 75
\item Lawrence, A., 1991, MNRAS, 252, 586
\item McCarthy, P.J., van Breugel, W.J.M., Spinrad, H., 
Djorgovski, S., 1987, ApJ, 321, L29
\item McCarthy, P.J., 1993, ARA\&A, 31, 639
\item McCarthy, P., Baum, S., Spinrad, H., 1996, ApJS, 99, 27.
\item Morganti, R., Ulrich, M.-H., Tadhunter, 1992, MNRAS, 254, 546
\item Morganti, R., Killean, N., Tadhunter, C.N., 1993, MNRAS,
263, 1023
\item Morganti, R., Oosterloo, T., Reynolds, J.E., Tadhunter, C.N., 
Migenes, V., 1997, MNRAS, 284, 541
\item Morganti, R., Tadhunter, C.N., Dickson, R.,  1997, A\&A, 326, 130
\item Osterbrock, D.E., Koski, A.T., Phillips, M.M., 1976, ApJ, 206, 898
\item Rawlings, S., Saunders, R., Eales, S.A., Mackay, C.D., 1989, 
MNRAS, 240, 701
\item Rawlings, S., Saunders, R., 1991, Nature, 349, 138
\item Robinson, A., Binette, L., Fosbury, R.E.A., Tadhunter, C.N., 1987,
MNRAS, 227, 97
\item Rudy, R.J., Tokunaga, A.T., 1982, ApJ, 256 L1
\item Saunders, R., Baldwin, J.E., Rawlings, S., Warner, P.J.,
Miller, L., 1989, MNRAS, 238, 777
\item Schmidt, M., 1965, ApJ, 141, 1
\item Siebert, J., Brinkmann, W., Morganti, R., Tadhunter, C.N., 
Danziger, I.J., Fosbury, R.A.E., di Serego Alighieri, S., 1996,
MNRAS, 279, 1331
\item Sutherland, R., Bicknell, G.V., Dopita, M.A., 1993, ApJ,
414, 510
\item Tadhunter, C.N., 1987, DPhil Thesis, University of Sussex.
\item Tadhunter, C.N., Robinson, A., Morganti, R., 1989. In: {\it ESO
Workshop on Extranuclear Activity in Galaxies}, eds Meurs, E.J.A., Fosbury,
R.A.E., ESO Conf. and Wokshop Proc. No.32, Garching, p293.
\item Tadhunter, C.N., 1991, MNRAS, 251, 46p
\item Tadhunter, C.N., Scarrott, S.M., Draper, P., Rolph, C., 1992, MNRAS, 256, 53p
\item Tadhunter, C.N., Morganti, R.M., di Serego Alighieri, S.,
Fosbury, R.A.E., Danziger, I.J., 1993, MNRAS, 263, 999
\item Tadhunter, C.N., Metz, S., Robinson, A., 1994, MNRAS, 268, 989
\item Tadhunter, C.N., Dickson, R., Villar-Martin, M., Morganti, 1997. In: {\it ESO/IAC Workshop on Quasar Hosts}, Clements et al. (eds), p.311
\item V\'eron-Cetty, M.-P., V\'eron, P., 1993. {\it A catalogue of quasars and
active galactic nuclei (6th edition)}, ESO Scientific Report No.13, ESO Publications.
\item Villar-Martin, M., Tadhunter, C., Morganti, R., Killeen, N.,
Clark, N., Axon, D., A\&A, 332, 479
\item Wall, J., Peacock, J., 1985, MNRAS, 216, 173
\item Young, S., Hough, J.H., Efstathiou, A., Wills, B.J., Bailey,
J.A., Ward, M.J., 1996, MNRAS, 279, L72
\item Zirbel, E.L., Baum, S.A., 1995, ApJ, 448, 521
 \end{description}
\newpage\noindent
{\bf Figure Captions}
\vglue 0.5cm\noindent
{\bf Figure 1.} The correlation between [OIII]$\lambda$5007 
emission line luminosity and total radio luminosity for the full $z < 0.7$
sample of Tadhunter \etal\ (1993) presented as 
a log-log plot (the units of the luminosities are erg s$^{-1}$). The
objects are classified accroding to their radio properties as follows:
filled circles --- Fanaroff Riley type II; open circles ---- Fanaroff Riley
type I; stars --- compact steep spectrum radio sources with
$D < 15$kpc; filled triangles ---- compact flat spectrum radio sources;
open diamonds --- uncertain radio classifications. The arrows indicate objects
for which we have only an upper limit in the  [OIII]$\lambda$5007 luminosity.
\vglue 0.3cm\noindent
{\bf Figure 2.} The correlations between the total radio luminosity and
(a) the [OIII]$\lambda$5007, (b) [OII]$\lambda$3727, and (c) H$\beta$ emission
line luminosities for the $0.1 < z < 0.7$ SSS subsample (see text for
definition). The correlations are presented as log-log plots (luminosity units:
erg s$^{-1}$). The symbols have the same definition as in Figure 1.
\vglue 0.3cm\noindent
{\bf Figure 3.} The ionization-sensitive [OII](3727)/[OIII](5007) emission
line ratio plotted against: (a) [OIII]$\lambda$5007 luminosity, 
(b)  [OII]$\lambda$3727 luminosity, (c)  H$\beta$ luminosity, and 
(d) total radio luminosity for the $0.1 < z < 0.7$ SSS sub-sample
(log-log plots). High redshift
($0.2 < z < 0.7$) and low redshift ($0.1 < z < 0.2$) objects are indicated
by filled and open symbols respectively; broad line objects are indicated
by circles and narrow line objects by squares. Note the separation
of high and low redshift objects.
\vglue 0.3cm\noindent
{\bf Figure 4.} The ionization-sensitive [OII](3727)/[OIII](5007) emission
line ratio plotted against [OIII]$\lambda$5007 emission line luminosity
for 3C radio galaxies (the symbols have the same meaning as Figure 3).
The data for the low redshift ($z < 0.2$) objects were taken from
Saunders \etal\ (1989), while the data for the higher redshift objects
were taken from Dickson (1997) and Jackson \& Rawlings (1997). 3C324, 3C368
and 3C265 are identified in the diagram to emphasise a possible link
between ionization state and radio/optical structure. 3C324 and 3C368 have
relatively compact radio sources and closely aligned radio
and UV structures, whereas 3C265 has a more extended radio source and a broader,
roughly biconical, emission line distribution.
\vglue 0.3cm\noindent
{\bf Figure 5.} The correlation between [OIII]$\lambda$5007 
emission line luminosity and total radio luminosity for the $0.1 < z < 0.7$
SSS sub-sample (log-log plot) with the broad line objects (including
BLRG and quasars) and narrow line line objects indicated by open and
filled circles respectively.
\vglue 0.3cm\noindent
{\bf Figure 6.} [OIII](5007)/H$\beta$ vs. [OII](3727)/[OIII](5007) diagnostic
diagram for the $0.1 < z < 0.7$ SSS sub-sample. The results of power-law
$F_{\nu} \propto \nu^{+\alpha}$ 
photoionization model calculations are shown as sequences in the
ionization parameter ($10^{-1} < U < 10^{-4}$, decreasing left to right)
for three different ionizing continuum shapes: $\alpha = -1.0$ (dotted line),
$\alpha = -1.5$ (dashed line),  and $\alpha = -2.0$ (dot-dashed line). Broad
line objects are indicated by squares, while narrow line objects are indicated
by circles.
\newpage\noindent
{\bf Table captions.}
\vglue 0.5cm\noindent
{\bf Table 1.} Basic data for the SSS sub-sample (see text for definition).
The optical classifications (column 3) are explained in the text, while
the radio classifications (column 4) 
are as follows: FRI --- Fanaroff Riley Class I;
FRII --- Fanaroff Riley Class II; CSS --- compact steep spectrum
radio source ($D < 15$kpc); and C/J --- core-jet (uncertain classifications
in brackets). The units of the total radio luminosities (column 6) and 
[OIII]$\lambda$5007 emission line luminosities (column 7) are log(erg s$^{-1}$).
The final two columns give, respectively, the [OII](3727)/[OIII](5007) and 
[OIII](5007)/H$\beta$ diagnostic emission line ratios. Most of the emission line
luminosities are based on the wide slit data presented
in Tadhunter \etal\ (1993), but
for cases  in which the data in Tadhunter \etal\ (1993) are incomplete, we have
used the deeper, but narrower slit, observations presented in Dickson (1997).
All the spectral classifications of the sources (column 3) have been updated
to take into account the new information in the deeper spectra (uncertain classifications
in brackets). Note that, although 0347+05 is listed as a WLRG, it could also be classified
as a BLRG since its spectrum shows broad permitted lines. The classification
of 1932-46 as a BLRG is based on the high resolution optical spectra
presented in Villar-Martin \etal\ (1997).
\vglue 0.3cm\noindent
{\bf Table 2.} Correlation analysis for the $0.1 < z < 0.7$ SSS sub-sample.
The first number given in each column is the Spearman's $\rho$
correlation coefficient, while the second number (in brackets) gives the
significance level of the correlation (\ie\ the percentage probability that the 
correlations could arise by chance). The analysis has been
performed separately for the full $0.1 < z < 0.7$ SSS sub-sample (column 2),
for the $0.1 < z < 0.7$ SSS sub-sample excluding the quasars (column 3), and
for the $0.1 < z < 0.7$ SSS sub-sample excluding quasars and WLRG
(column 4).
\vglue 0.3cm\noindent
{\bf Table 3.} Median ionization state ([OII](3727)/[OIII](5007)),
total radio luminosities, and [OIII]$\lambda$5007 emission line 
luminosities for the SSS subsample in different redshift ranges. 
Note that there is no evidence
for a systematic increase in ionization state with redshift and radio power.

\newpage

% Fig. 1
\centerline{\psfig{figure=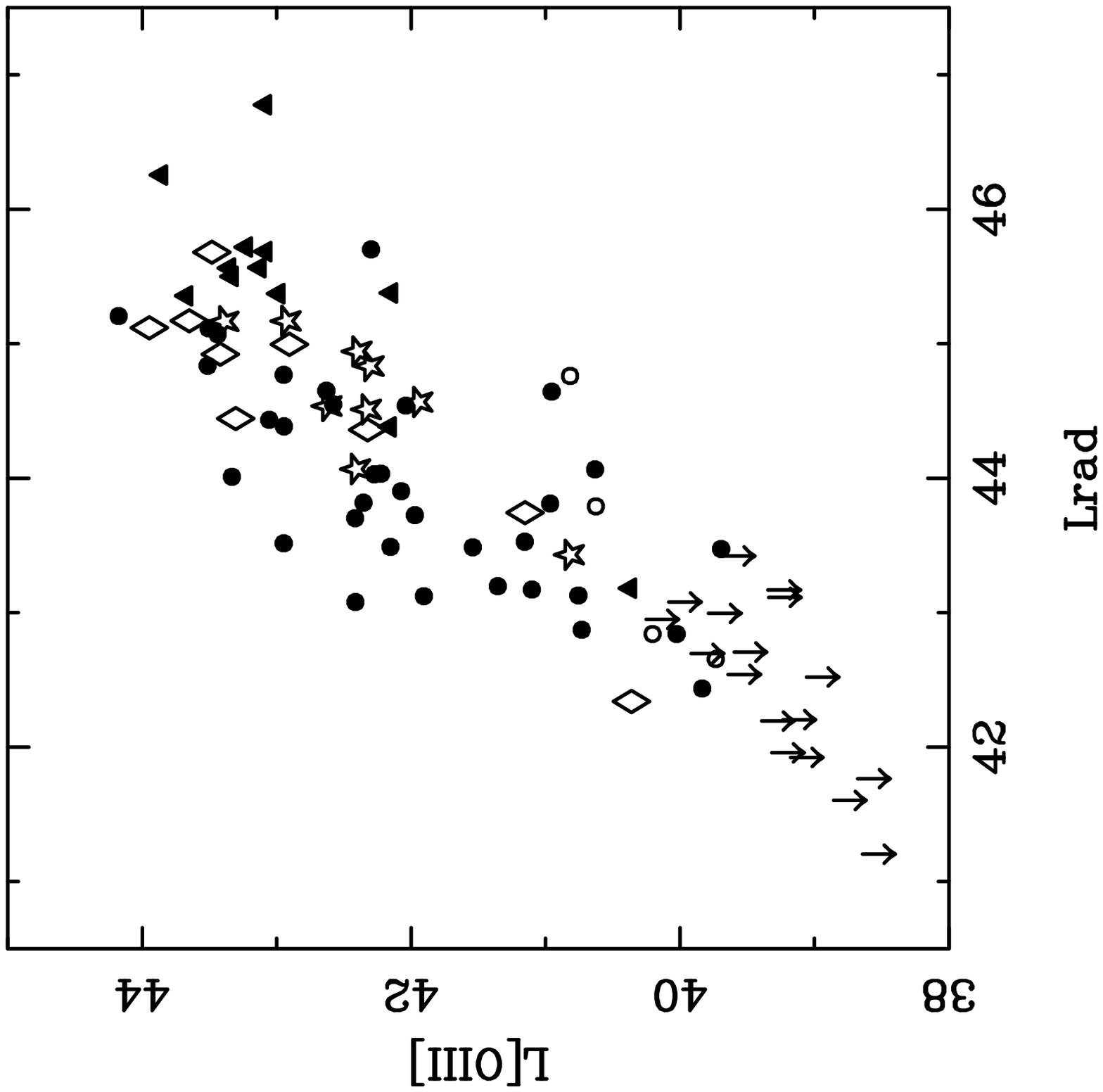,width=12cm,angle=0}}
%{\bf Fig.1}

\newpage

% Fig. 2
\centerline{\psfig{figure=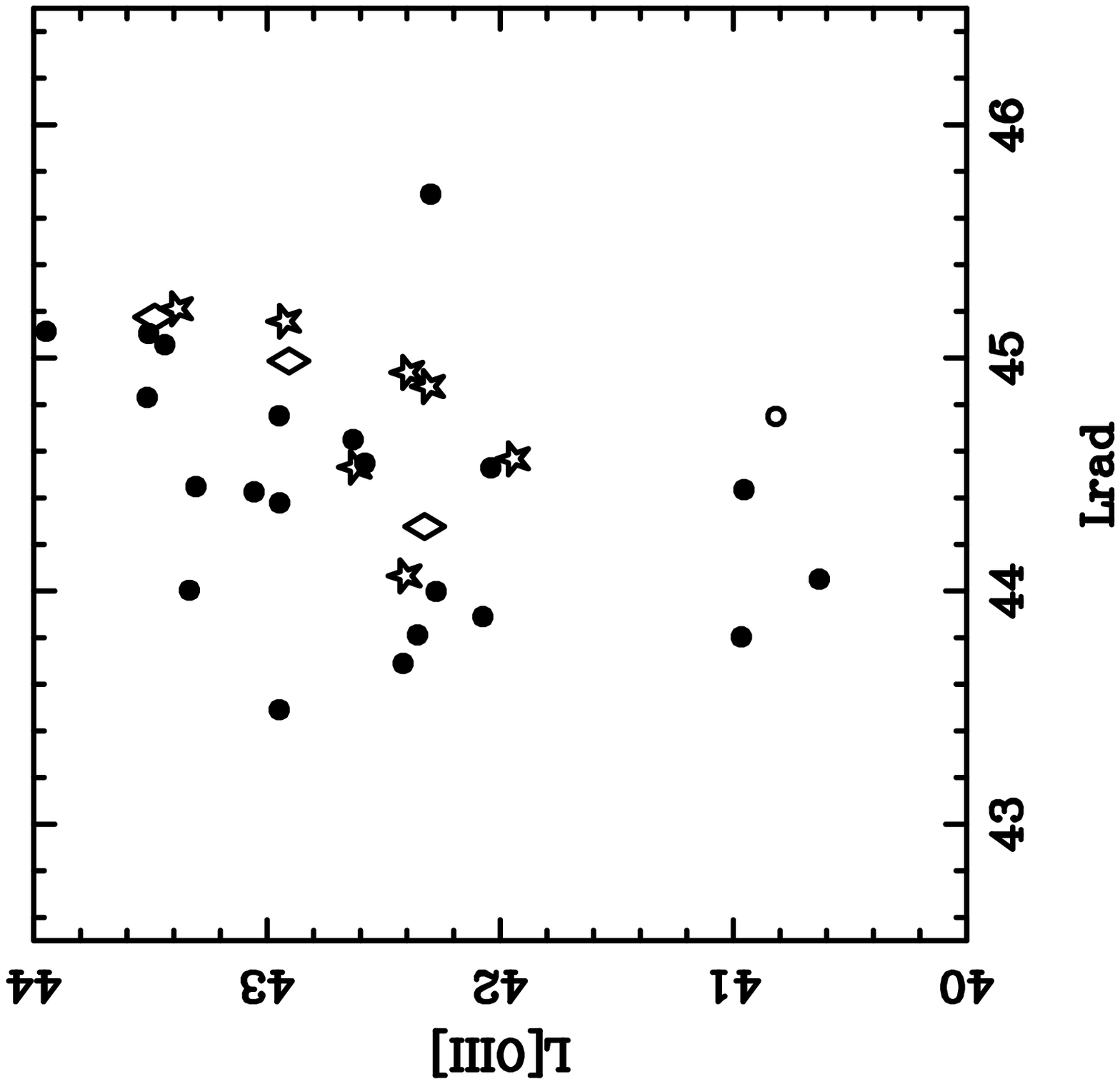,width=7cm,angle=0}}
%{\bf Fig.2a}
\newpage
\centerline{\psfig{figure=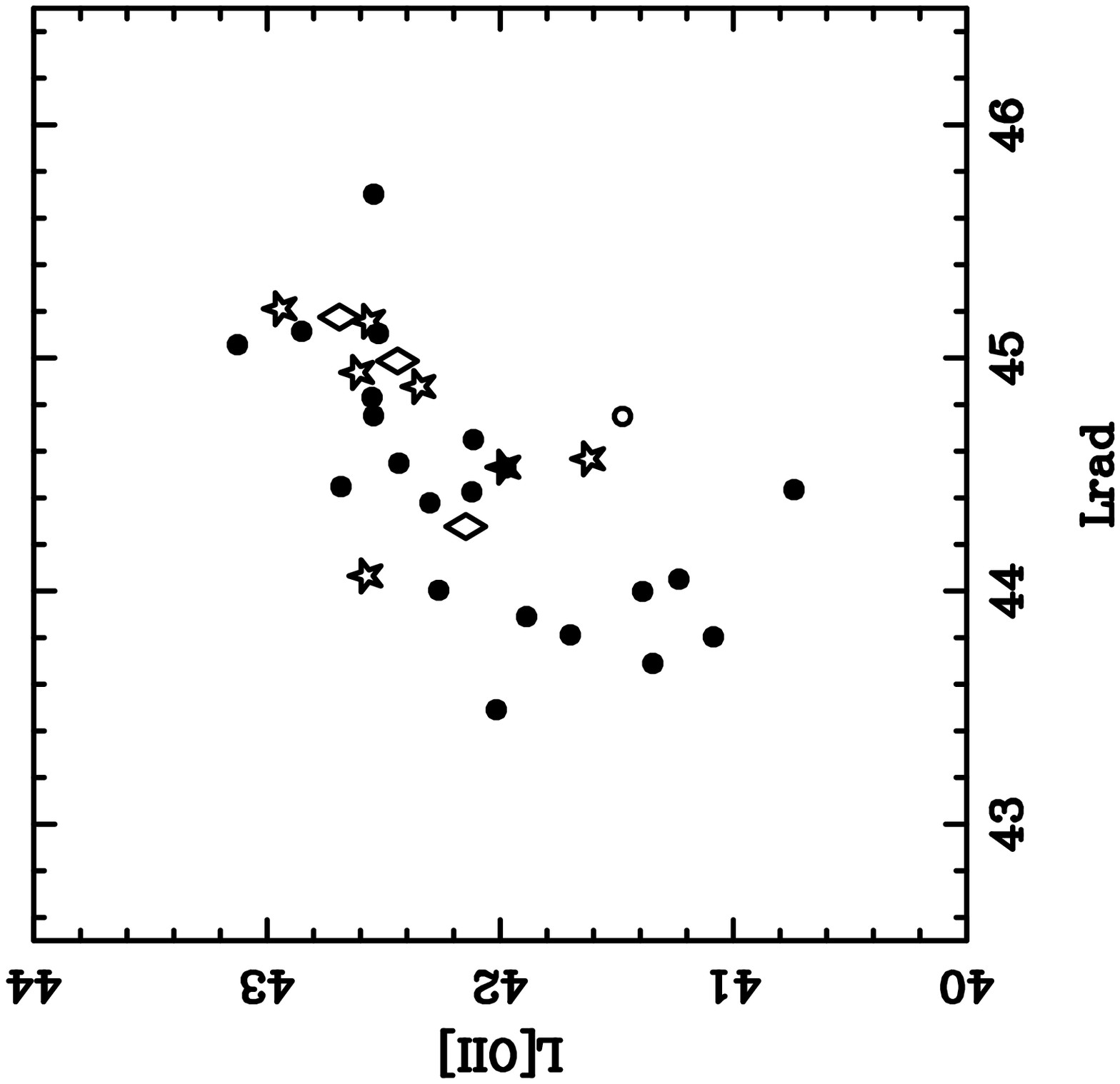,width=7cm,angle=0}}
%{\bf Fig.2b}
\newpage
\centerline{\psfig{figure=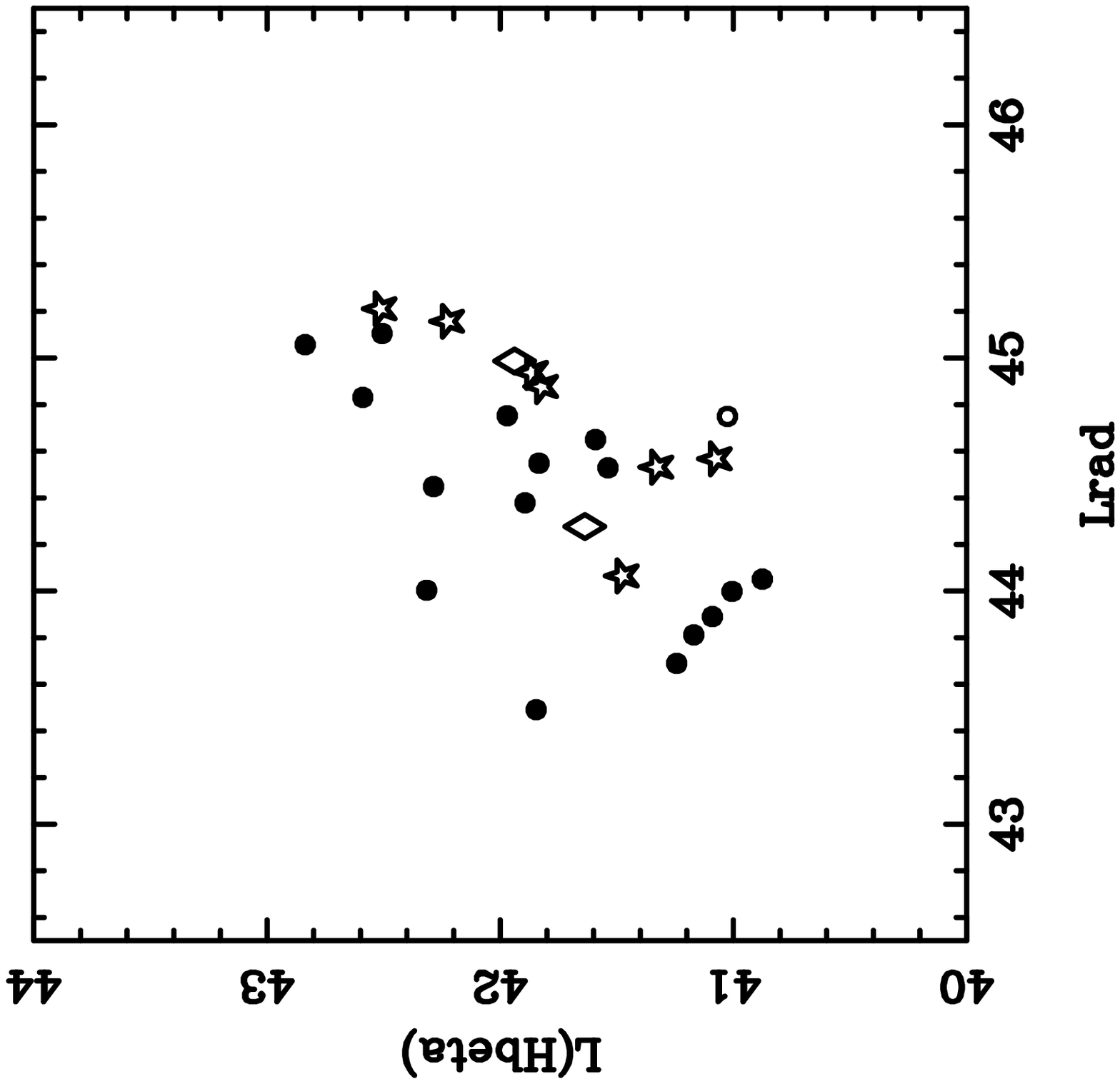,width=7cm,angle=0}}
%{\bf Fig.2c}
\newpage

\centerline{\psfig{figure=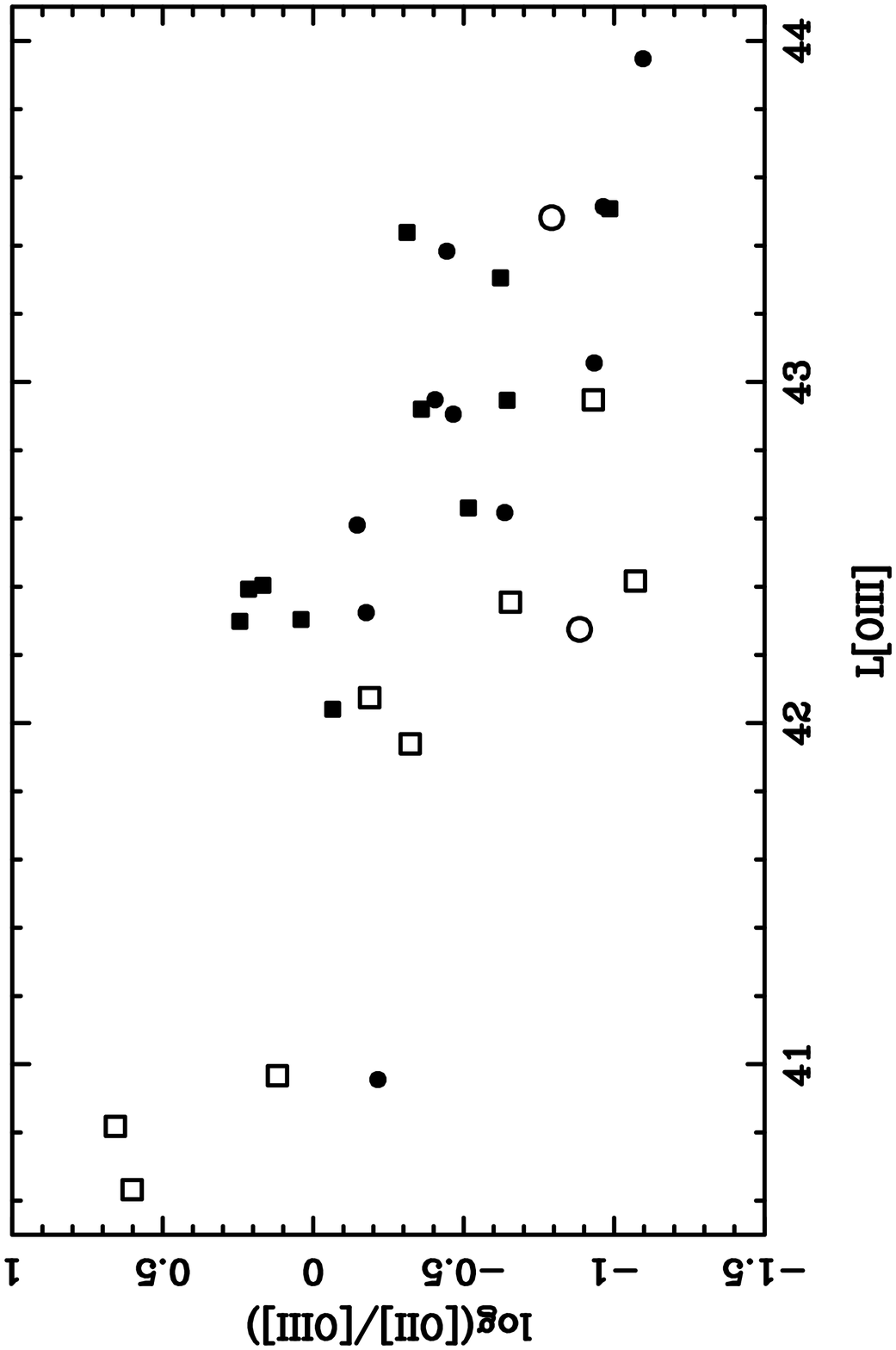,width=7cm,angle=0}}
%{\bf Fig.3a}
\newpage
\centerline{\psfig{figure=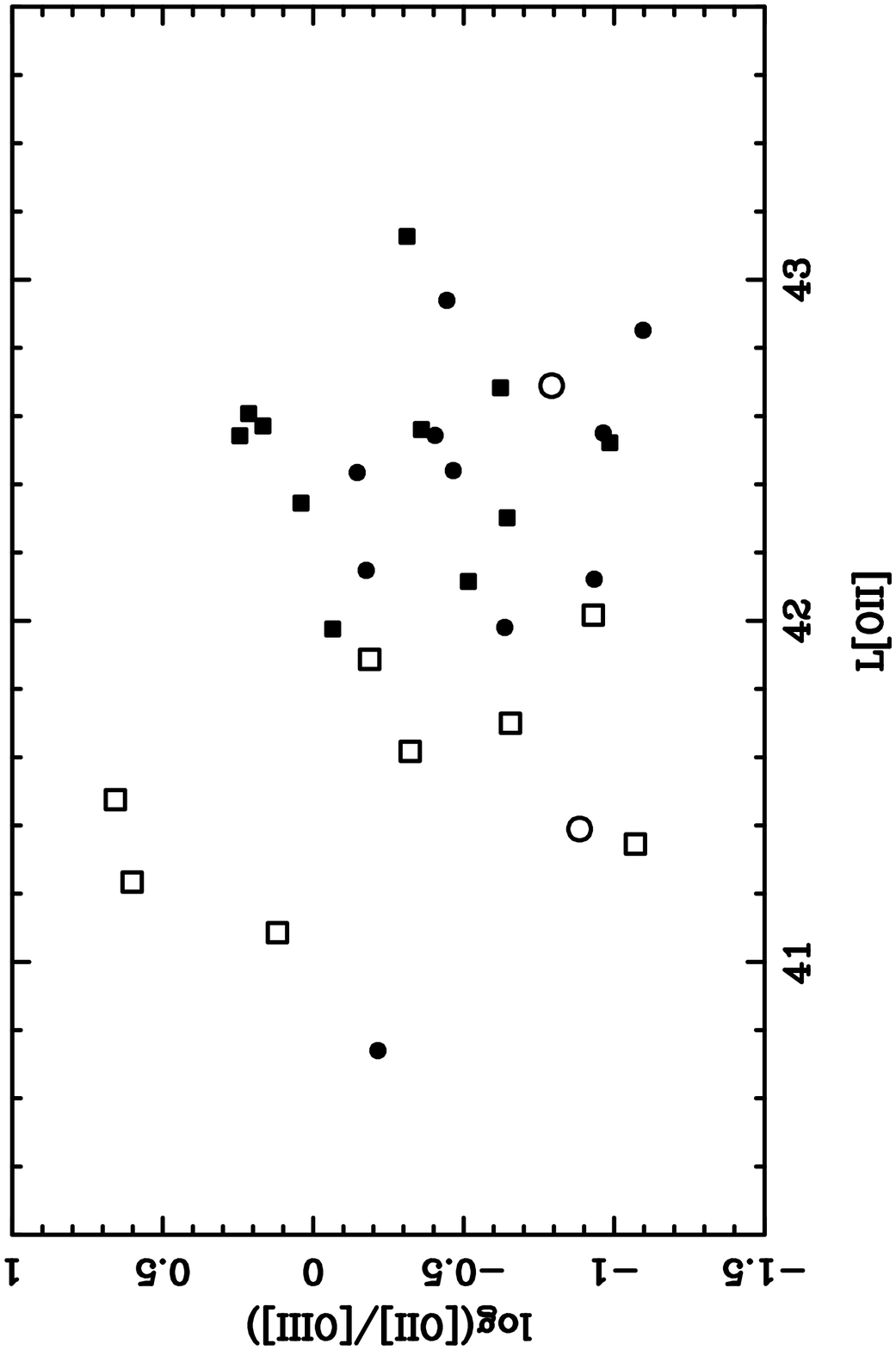,width=7cm,angle=0}}
%{\bf Fig.3b}
\newpage
\centerline{\psfig{figure=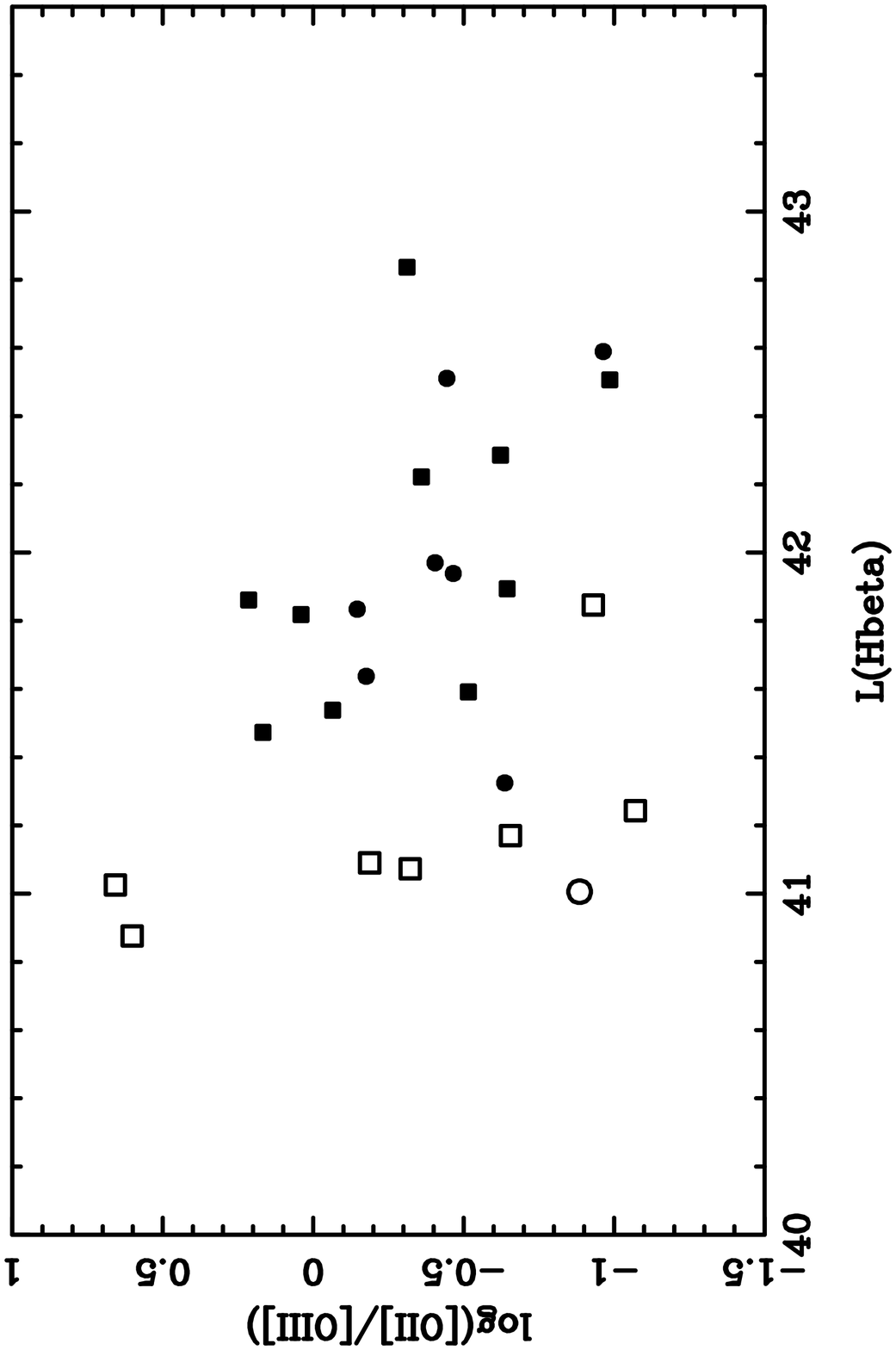,width=7cm,angle=0}}
%{\bf Fig.3c}
\newpage
\centerline{\psfig{figure=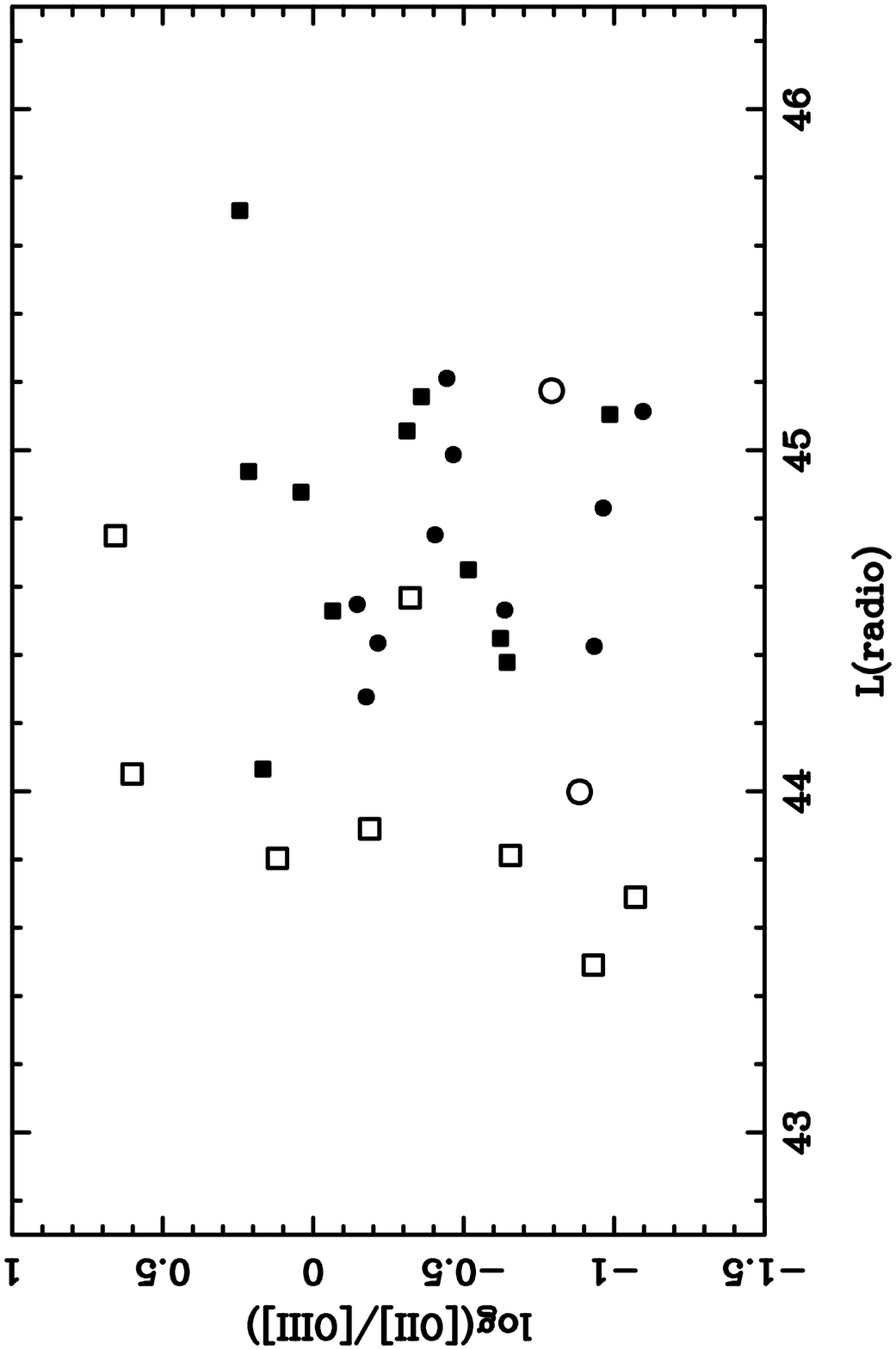,width=7cm,angle=0}}
%{\bf Fig.3d}
\newpage
\centerline{\psfig{figure=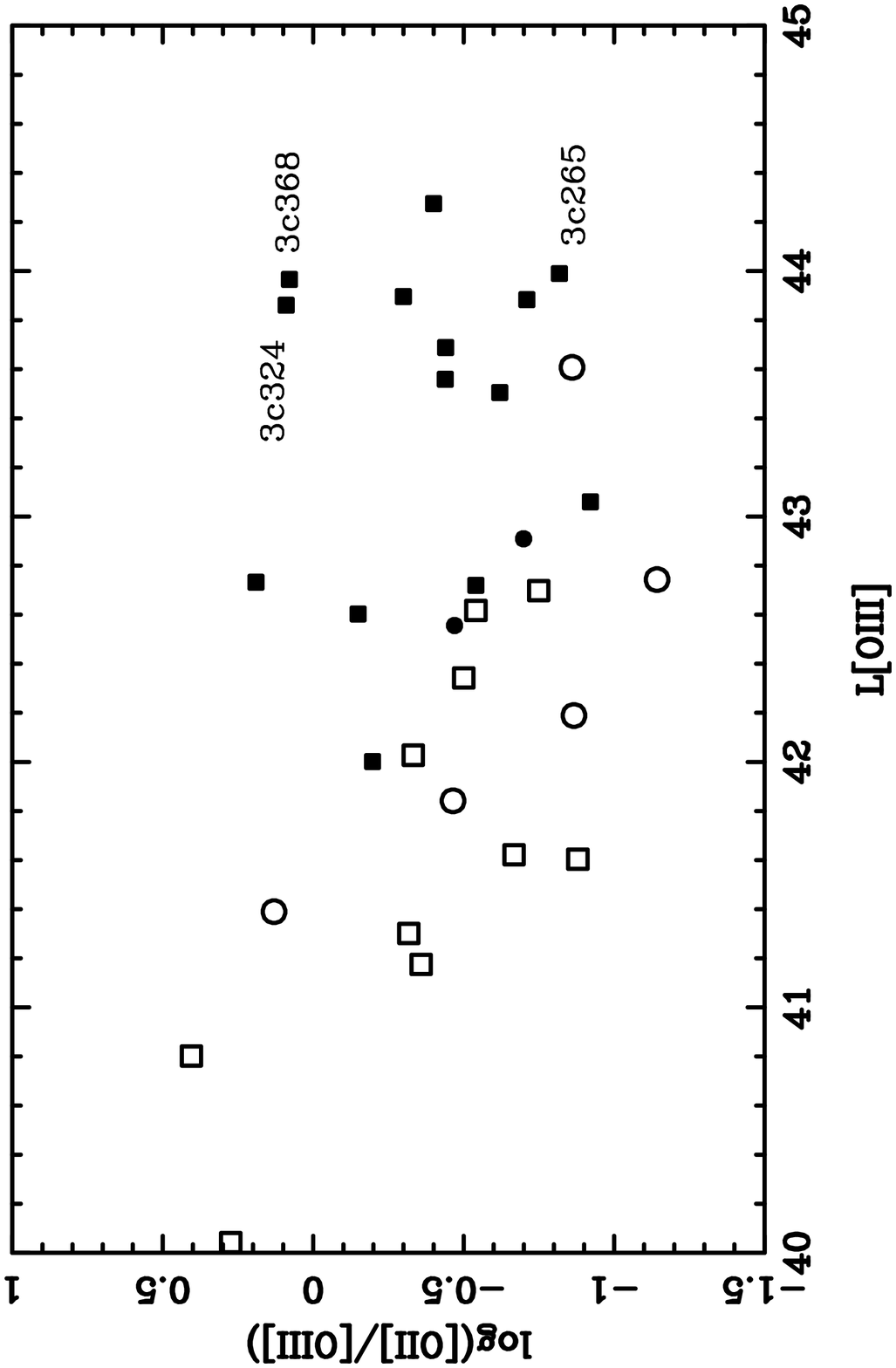,width=7cm,angle=0}}
%{\bf Fig.4}
\newpage
\centerline{\psfig{figure=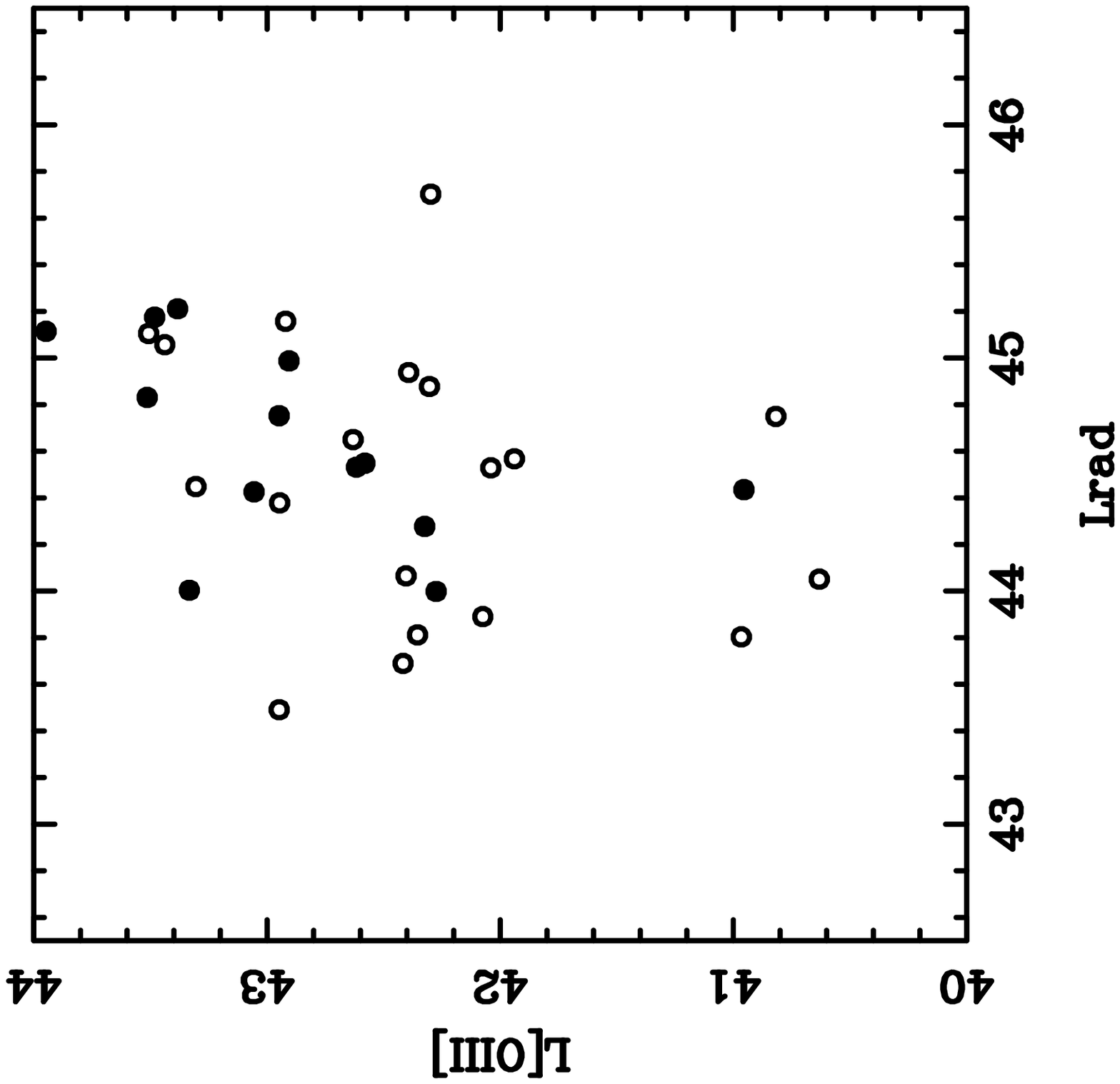,width=7cm,angle=0}}
%{\bf Fig.5}
\newpage
\begin{figure}
\centerline{\psfig{figure=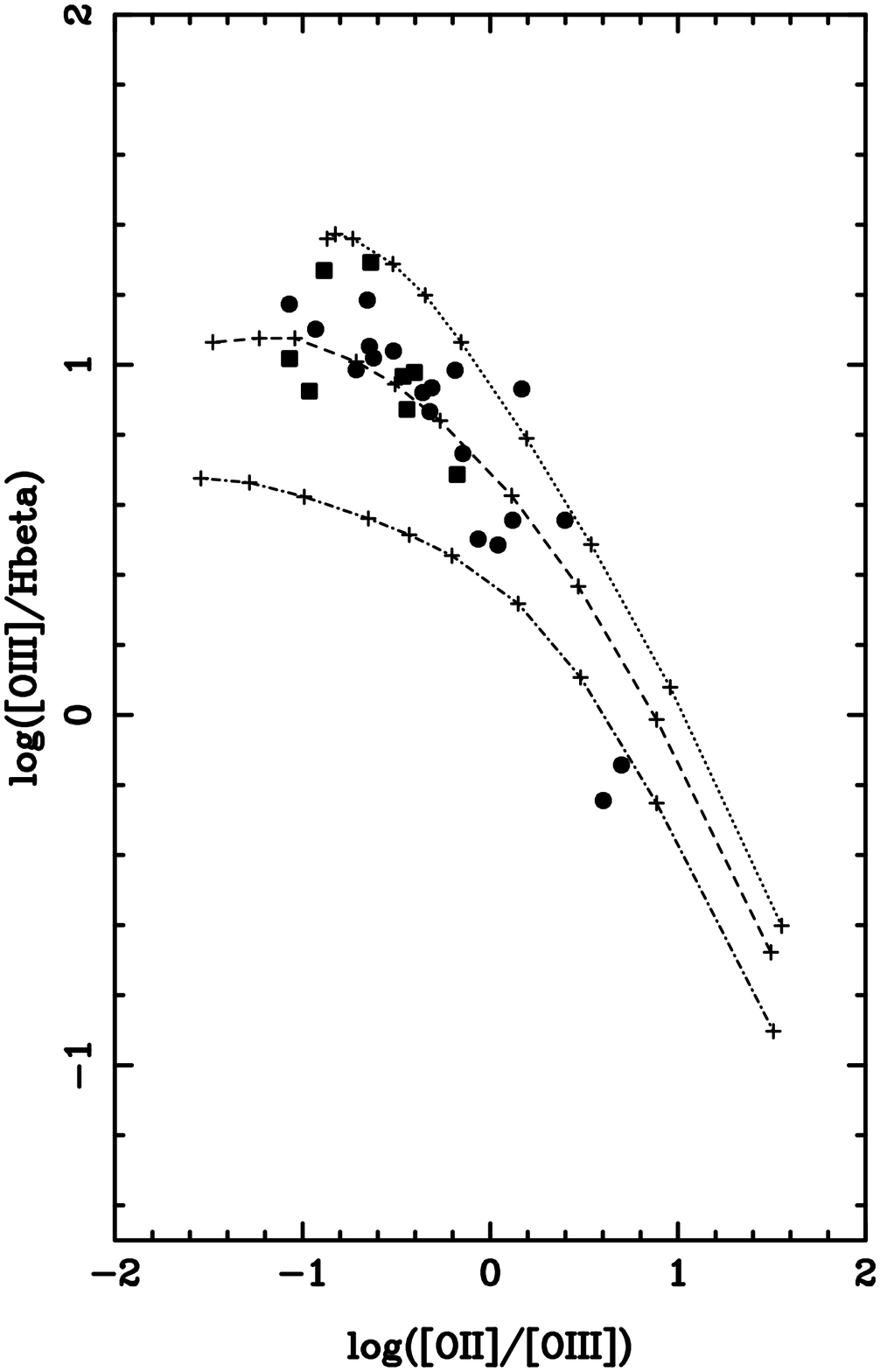,width=5cm,angle=0}}
%{\bf Fig.6}
\end{figure}

\newpage

\begin{table}
\begin{tabular}{llllllrll} \hline \hline
Object & & & &z &L$_{rad}$ &L$_{[OIII]}$ &[OIII]/[OII] &[OIII]/H$_{\beta}$
\\  \hline
0023--26 & &NLRG &CSS &0.322  &44.88  &42.30   &0.91   &3.06 \\
0034--01  & &WLRG &FRII &0.073  &43.07 &40.76  &---   &3.69 \\
0035--02  &3C17 &BLRG &(FRII)    &0.220  &44.28  &42.32   &1.50   &4.86 \\
0038+09  &3C18 &BLRG &FRII &0.188  &44.00 &42.27   &7.68  &18.58 \\
0039--44  & &NLRG &FRII &0.346  &44.45  &43.31   &4.19  &10.46 \\
0043--42  & &WLRG &FRII &0.116  &43.80  &40.97   &0.76   &--- \\
0055--01  &3C29 &WLRG &FRI &0.045  &42.83  &40.20  &---   &1.65 \\
0105--16  &3C32 &NLRG &FRII &0.400  &44.65  &42.63   &3.28  &10.95 \\
0117--15  &3C38 &NLRG &FRII &0.565  &45.06  &43.44   &2.05   &4.00 \\
0123--01  &3C40 &WLRG &FRII &0.018  &41.95  &$<$39.20  &--- &--- \\
0131--36  &NGC612 &WLRG &FRII &0.030  &42.84  &40.03  &--- &3.14 \\
0213--13  &3C62 &NLRG &FRII &0.147  &43.81  &42.35   &4.52  &15.30 \\
0235--19  & &BLRG &FRII &0.620  &45.10  &43.51   &9.68  &10.03 \\
0252--71  & &NLRG &CSS &0.566  &45.16  &42.92   &2.29   &5.00 \\
0255+05  &3C75 &WLRG &FRI &0.023  &42.19  &$<$39.27 &---  &--- \\
0305+03  &3C78 &WLRG &FRI &0.029  &42.62  &$<$39.48  &--- &--- \\
0320--37  &Fornax A &WLRG &FRI &0.005  &42.52  &$<$38.94  &--- &--- \\
0325+02  &3C88 &WLRG &FRII &0.030  &42.40  &39.84  &--- &--- \\
0347+05  & &BLRG &FRII &0.339 &44.43 &40.95 &1.64 &--- \\
0349--27  & &NLRG &FRII &0.066  &43.19  &41.36   &0.59   &6.90 \\
0404+03  &3C105 &NLRG &FRII &0.089  &43.53  &41.16  &---  &15.92 \\
0409--75  & &NLRG &FRII &0.693  &45.70  &42.30   &0.57  &--- \\
0428--53  & &WLRG &FRI &0.038  &43.16  &$<$39.22 &--- &--- \\  
0442--28  & &NLRG &FRII &0.147  &43.89  &42.07   &1.54   &9.65 \\
0453--20  & &WLRG &FRI &0.035  &42.54  &$<$39.52  &---  &--- \\
0518--45  &Pictor A &BLRG &FRII &0.035  &43.46  &41.54   &1.77   &4.21 \\
0521--36  & &BLRG &C/J &0.055  &43.69  &41.15   &3.13   &5.47 \\
0625--53  & &WLRG &(FRII) &0.054  &42.97  &$<$39.67  &---  &--- \\
0625--35  & &WLRG &FRI &0.055  &42.97  &$<$39.96  &--- &--- \\
0806--10  & &NLRG &FRII &0.110  &43.49  &42.95   &8.53  &12.64 \\
0859--25  & &NLRG &FRII &0.305  &44.53  &42.04   &1.16   &3.18 \\
0915--11  &Hydra A &WLRG &FRI &0.054  &43.78  &40.63   &0.15   &0.75 \\
0945+07  &3C227 &BLRG &FRII &0.086  &43.47  &42.15  &12.67   &9.01 \\
1136--13  & &Q &FRII &0.554  &45.11  &43.95  &12.48  &--- \\
1151--34  & &Q &CSS &0.258  &44.53  &42.62  &4.33  &19.61 \\
1216+06  &3C270 &WLRG &FRI &0.006  &41.75  &$<$38.56 &---  &--- \\
1226+02  &3C273 &Q &C/J &0.158  &45.17  &43.48   &---  &--- \\
1246--41  &NGC4696 &WLRG &FRI &0.009  &41.20  &$<$38.52  &--- &--- \\
1251--12  &3C278 &WLRG &FRI &0.015  &41.92  &$<$39.06  &---  &--- \\
1306--09  & &NLRG &CSS &0.464  &44.94  &42.39   &0.61   &3.40 \\
1318--43  &NGC5090 &WLRG &FRI &0.011  &41.62  &$<$38.74  &--- &--- \\
1333--33  &IC4296 &WLRG &FRI &0.013  &41.96  &$<$39.11  &--- &--- \\
1355--41  & &Q &FRII &0.313  &44.43  &43.06   &8.59  &--- \\
\hline
\hline
\end{tabular}
\end{table}

\newpage\noindent

\begin{table}
\begin{tabular}{lllllrrrr} \hline \hline
Object & & & &z &L$_{rad}$ &L$_{[OIII]}$ &[OIII]/[OII] &[OIII]/H$_{\beta}$
\\ \hline
1547--79  & &BLRG &FRII &0.483  &44.83  &43.51   &9.20   &8.41 \\
1559+02  &3C327 &NLRG &FRII &0.104  &43.69  &42.42  &11.78  &14.90 \\
1602+01  &3C327.1 &NLRG &FRII &0.462  &44.75  &42.95   &2.54   &9.51 \\
1637-77  & &WLRG &FRII &0.041  &42.83  &40.73   &1.06   &9.61 \\
1648+05  &Hercules A &WLRG &FRI/FRII &0.154  &44.75  &40.82   &0.22   &0.62 \\
1717-00  &3C353 &WLRG &FRII &0.031  &43.46  &39.69  &---   &0.92 \\
1733-56  &  &BLRG &FRII &0.098 &43.42  &41.97   &1.39   &5.72 \\
1814-63  & &NLRG &CSS &0.063  &43.42  &40.81  &--- &--- \\
1932-46  & &BLRG &FRII &0.231  &44.55  &42.58   &1.40   &5.58 \\
1934-63  & &NLRG &CSS &0.183  &44.57 &41.94   &2.10   &7.35 \\
1938-15  & &BLRG &FRII &0.452  &44.99  &42.91   &2.92   &9.27 \\
1949+02  &3C403 &NLRG &FRII &0.059  &43.12  &41.91   &8.41  &20.10 \\
1954-55  & &WLRG &FRI &0.060  &43.05  &$<$39.22  &--- &--- \\
2058-28  & &WLRG &FRI &0.038  &42.63  &39.73  &---   &1.03 \\
2104-25  & &WLRG &FRII &0.037  &42.71  &40.13 &--- &--- \\
2135-14  & &Q  &FRII &0.200 &44.00 &43.33 &11.74 &10.42 \\
2135-20  & &BLRG &CSS &0.635  &45.21  &43.38 &2.78 &7.46 \\
2152-69  & &BLRG &FRII &0.027 &43.16  &41.10 &2.47 &7.42 \\
2211-17  &3C444 &WLRG &FRII &0.153  &44.05  &40.63 &0.25  &0.57 \\
2221-02  &3C445 &BLRG &FRII &0.057  &43.07  &42.41  &14.10  &12.88 \\
2250-41  & &NLRG &FRII &0.310  &44.38  &42.95   &4.41  &11.29 \\
2314+03  &3C459  &NLRG &(CSS) &0.220  &44.07  &42.40   &0.68   &8.53 \\
2356-61  & &NLRG &FRII &0.096  &44.02  &42.22   &2.65  &12.62 \\
\hline
\hline
\end{tabular}
\end{table}
\newpage

\begin{figure}
\centerline{\psfig{figure=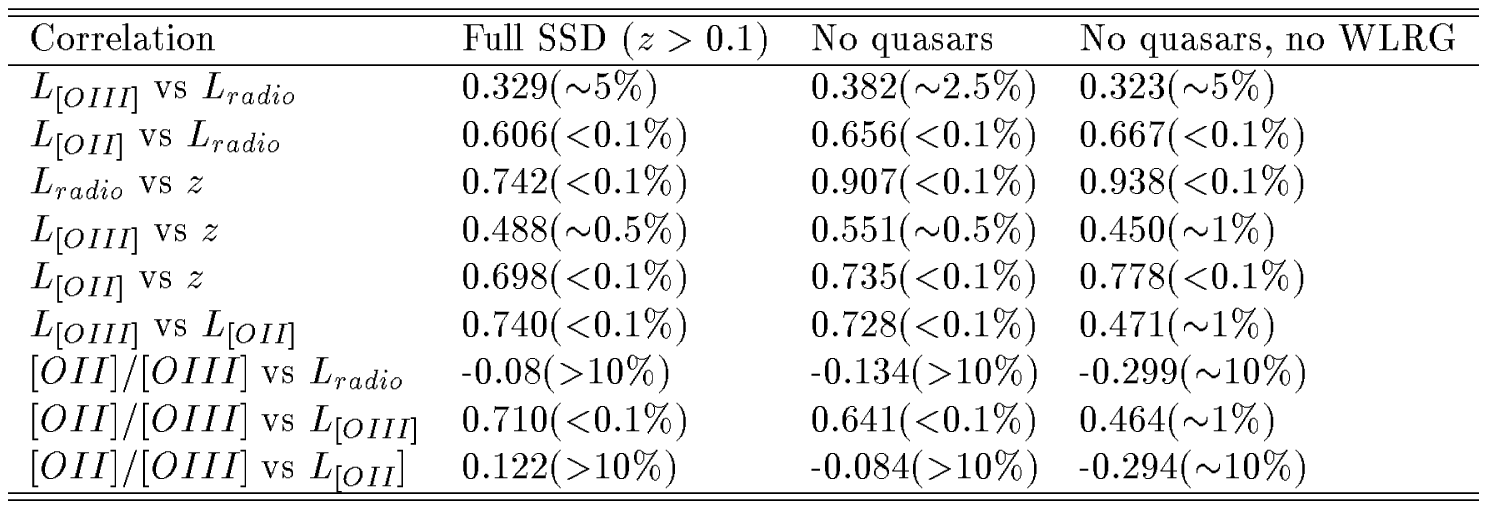,width=20cm,angle=0}}
\end{figure}

\newpage
\begin{figure}
\centerline{\psfig{figure=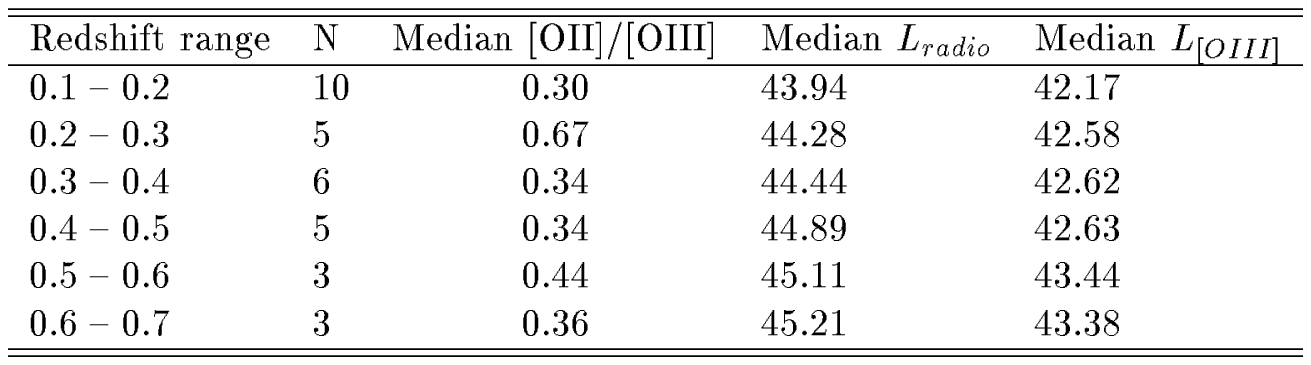,width=20cm,angle=0}}
\end{figure}

\end{document}